\documentclass[useAMS,usenatbib]{mn2e}

\usepackage{amsmath}
\usepackage{amssymb}
\usepackage{graphicx}
\usepackage{aas_macros}

\title[SILCC: Dynamics of the SN-driven ISM]{The SILCC (SImulating the LifeCycle of molecular Clouds) project - II. Dynamical evolution of the supernova-driven ISM and the launching of outflows}
\author[Girichidis et al.]
       {Philipp~Girichidis$^{1}$\thanks{email: \texttt{philipp@girichidis.com}},
         Stefanie~Walch$^{2}$,
         Thorsten~Naab$^{1}$,
         Andrea~Gatto$^{1}$,
         \newauthor
         Richard~W\"{u}nsch$^{3}$,
         Simon~C.~O.~Glover$^{4}$,
         Ralf~S.~Klessen$^{4}$,
         Paul~C.~Clark$^{5}$,
         \newauthor
         Thomas~Peters$^{1,6}$,
         Dominik~Derigs$^{2}$,
         Christian~Baczynski$^{4}$
         ~\\
         $^1$Max-Planck-Institut f\"{u}r Astrophysik, Karl-Schwarzschild-Str. 1, 85741 Garching, Germany\\
         $^2$Physikalisches Institut, Universit\"{a}t zu K\"{o}ln, Z\"{u}lpicher Str. 77, 50937 K\"{o}ln, Germany\\
         $^3$Astronomical Institute, Academy of Sciences of the Czech Republic, Bocni II 1401, 141 31 Prague, Czech Republic\\
         $^4$Universit\"{a}t Heidelberg, Zentrum f\"{u}r Astronomie, Institut f\"{u}r Theoretische Astrophysik, Albert-Ueberle-Str. 2, 69120 Heidelberg, Germany\\
         $^5$School of Physics \& Astronomy, Cardiff University, 5 The Parade, Cardiff CF24 3AA, Wales, UK\\
         $^6$Institut f\"{u}r Computergest\"{u}tzte Wissenschaften, Universit\"{a}t Z\"{u}rich, Winterthurerstr. 190, CH-8057 Z\"{u}rich, Switzerland}

\newcommand{\citetWuenschEtAl}{W\"{u}nsch et al. (in preparation)}
\newcommand{\citepWuenschEtAl}{(W\"{u}nsch et al., in preparation)}

\begin{document}

\maketitle

\begin{abstract}
  The SILCC project (SImulating the Life-Cycle of molecular Clouds) aims at a more self-consistent understanding of the interstellar medium (ISM) on small scales and its link to galaxy evolution. We present three-dimensional (magneto)hydrodynamic simulations of the ISM in a vertically stratified box including self-gravity, an external potential due to the stellar component of the galactic disc, and stellar feedback in the form of an interstellar radiation field and supernovae (SNe). The cooling of the gas is based on a chemical network that follows the abundances of H$^+$, H, H$_2$, C$^+$, and CO and takes shielding into account consistently. We vary the SN feedback by comparing different SN rates, clustering and different positioning, in particular SNe in density peaks and at random positions, which has a major impact on the dynamics. Only for random SN positions the energy is injected in sufficiently low-density environments to reduce energy losses and enhance the effective kinetic coupling of the SNe with the gas. This leads to more realistic velocity dispersions ($\sigma_\mathrm{HI}\approx0.8\sigma_{300-8000\,\mathrm{K}}\sim10-20\,\mathrm{km}\,\mathrm{s}^{-1}$, $\sigma_\mathrm{H\alpha}\approx0.6\sigma_{8000-3\times10^5\,\mathrm{K}}\sim20-30\,\mathrm{km}\,\mathrm{s}^{-1}$), and strong outflows with mass loading factors (ratio of outflow to star formation rate) of up to $10$ even for solar neighbourhood conditions. Clustered SNe abet the onset of outflows compared to individual SNe but do not influence the net outflow rate. The outflows do not contain any molecular gas and are mainly composed of atomic hydrogen. The bulk of the outflowing mass is dense ($\rho\sim10^{-25}-10^{-24}\,\mathrm{g\,cm}^{-3}$) and slow ($v\sim20-40\,\mathrm{km}\,\mathrm{s}^{-1}$) but there is a high-velocity tail of up to $v\sim500\,\mathrm{km}\,\mathrm{s}^{-1}$ with $\rho\sim10^{-28}-10^{-27}\,\mathrm{g\,cm}^{-3}$.
\end{abstract}


%%%%%%%%%%%%%%%%%%%%%%%
\section{Introduction}%
%%%%%%%%%%%%%%%%%%%%%%%
\label{sec:introduction}
Star formation and the resulting stellar feedback, together with the thermal, chemical and dynamical evolution of the galactic disc, drive the `matter cycle' in the interstellar medium (ISM). It describes the way in which gas is cycled from a hot, diffuse ionized phase into colder, denser phases, and ultimately into molecular gas and back. Star formation occurs within this molecular gas, and the resulting stellar feedback both heats the ISM and returns stellar material to the ISM. At the same time the cold gas is heated and dispersed, while the momentum deposited by outflows and winds drives turbulent motions in the gas. The strongest feedback is expected to stem from the thermal and dynamical impact of supernovae (SNe), which play a major role in the overall dynamics of the ISM \citep{McKeeOstriker1977, MacLowKlessen2004, KlessenGlover2014}.

The effective impact of SN feedback strongly depends on the environment in which the SNe explode \citep{Taylor1950,Sedov1959,McKeeOstriker1977,CowieMcKeeOstriker1981,OstrikerMcKee1988,CioffiMcKeeBertschinger1988,SlavinCox1992,DwarkadasGruszko2012,RogersPittard2013,WalchNaab2015,KimOstriker2015,GattoEtAl2015,IffrigHennebelle2015,MartizziFaucherGiguereQuataert2015,LiEtAl2015}. SNe in dense regions violently interact with their immediate surroundings, potentially destroying molecular gas in their vicinity. Efficient cooling can convert most of the energy into radiation that escapes the SN site and reduces the region of dynamical influence. The shells of SNe exploding in low-density environments can expand to larger distances and reshape the ISM on larger scales.

Observations of SN remnants in the Milky Way indicate that $\sim20\%$ of the SNe are of type~Ia. Their explosion sites follow an exponential distribution centred around the midplane with a scale height of $\sim300\,\mathrm{pc}$. The remaining SNe are of type~II, again with an exponential distribution around the midplane but significantly smaller scale height of $\sim50\,\mathrm{pc}$ \citep{TammannLoefflerSchroeder1994}. Stars typically form in clusters over a wide range of masses and spatial scales \citet[e.g.][]{LadaLada2003}. Most of the massive stars \citep[$\gtrsim3/4$ of the O stars,][]{deWitEtAl2004} are also clustered, and hence so are most of the type~II SNe. However, a fraction of the type~II SNe are distributed, due to either isolated star formation \citep{KennicuttEdgarHodge1989,McKeeWilliams1997,ClarkeOey2002,SchilbachRoeser2008} or runaway stars \citep{Blaauw1961,GiesBolton1986,GvaramadzeBomans2008,EldridgeLangerTout2011,PeretsSubr2012}.

Stellar feedback is also expected to drive galactic fountains, outflows and winds \citep[see, e.g.][]{ChevalierClegg1985,MurrayQuataertThompson2005}. Observations reveal that gas is ejected from the disc at velocities of a $\mathrm{few}\,100\,\mathrm{km}\,\mathrm{s}^{-1}$ with mass loading factors (ratio of outflowing to star-forming gas) of order unity and above \citep[see, e.g.][]{ChenEtAl2010,MartinEtAl2012,NewmanEtAl2012bShort,MartinEtAl2013}. Simulations of galaxy evolution in a cosmological framework support the dynamical impact of fountains and outflows and emphasize their importance for metal enrichment \citep{OppenheimerDave2008,OppenheimerEtAl2010,StinsonEtAl2010,HopkinsQuataertMurray2012,HirschmannEtAl2013}.

The properties and the dynamical evolution of the ISM have been investigated with numerical simulations of varying complexity in terms of the physical processes considered. A classical setup to numerically study ISM properties are stratified boxes covering a statistically significant volume of the ISM with typical sizes ($x\times y\times\pm z$) of $x=y\sim0.5-1\,\mathrm{kpc}$, up to $z\sim\pm20\,\mathrm{kpc}$ \citep{deAvillez2000,deAvillezBerry2001,deAvillezBreitschwerdt2004,deAvillezBreitschwerdt2005, JoungMacLow2006, PiontekOstriker2007, JoungMacLowBryan2009, KoyamaOstriker2009a,KoyamaOstriker2009b, KimKimOstriker2011, ShettyOstriker2012, HillEtAl2012, CreaseyTheunsBower2013, GentEtAl2013a,GentEtAl2013b, HennebelleIffrig2014, SILCC1}. In the ISM it is important to distinguish between the scales of the dynamical evolution of the disc, namely the local turbulent and thermal structures, and the scales of fountains, outflows, and chimneys that determine the large scale gas cycle. For the small scale dynamics in the disc, most studies find converged velocity dispersions on time-scales of several tens of $\mathrm{Myr}$. The coupling of small scale motions to large scale outflows takes an order of magnitude longer. Early numerical work on the SN-driven ISM by \citet{deAvillezBerry2001} report the formation of collimated chimneys of outflowing gas after $\sim100\,\mathrm{Myr}$. Both \citet{deAvillezBreitschwerdt2004,deAvillezBreitschwerdt2005} and \citet{HillEtAl2012} emphasize that time-scales of a few hundred $\mathrm{Myr}$ are needed for a fountain cycle to establish a global dynamical equilibrium and a converged vertical profile of the stratified disc. \citet{PiontekOstriker2007} use galactic shear driven magneto-rotational instability instead of SNe to drive turbulent motions and establish a dynamical equilibrium in the vertical stratification, however, with perceptibly lower velocity dispersions in the gas. In their models it takes more than a Gyr to establish dynamical equilibrium. \citet{CreaseyTheunsBower2013} quantitatively connect the outflow rates to the gas content in the disc and the amount of stellar feedback which is proportional to the star formation efficiency. They find mass loading factors ($\eta = \dot{M}_\mathrm{outflow}/\dot{M}_\mathrm{SFR}$) ranging from $\eta\ll0.1-4$ that have a weak dependence on the star formation rate but significant dependence on the gas fraction and surface density of the disc.

We also note that the Galactic ISM is permeated by magnetic fields. On galactic scales the field is expected to be generated by the mean-field dynamo with field strengths of $\sim\,1-10\,\mu\mathrm{G}$ \citep[see e.g.][]{Beck2001,Beck2009,CrutcherEtAl2010}. The field is dominated by the toroidal component. Local tangled components are of similar strengths, however, in dense molecular clouds fields up to several mG have been measured \citep[see e.g.][]{CrutcherEtAl2010, Crutcher2012}.

Two different aspects of the ISM have not been investigated in previous studies of stratified box models, namely the positioning and clustering of SNe in concert with a more sophisticated thermodynamic treatment using a chemical network and including shielding and self-shielding of the gas. Within the SILCC project (SImulating the Life-Cycle of molecular Clouds\footnote{For movies of the simulations and download of selected simulation data see the SILCC website: \texttt{www.astro.uni-koeln.de/silcc}}), starting with \citet{SILCC1}, we aim to advance our understanding of the physical processes in the ISM by investigating the impact of differently placed SNe as individual explosions as well as clustered explosions. \citet{SILCC1} find that only simulations with random or clustered SN positioning are in agreement with observations. In those models molecular hydrogen contributes $40\%-60\%$ to the total mass, whereas the other half of the mass is in atomic hydrogen. This paper focuses on the dynamical evolution of the gas in the disc. This includes the velocity dispersion, the midplane pressures and the resulting onset of outflows.

%%%%%%%%%%%%%%%%%%%%%%%%%%%%%%%%%%%%%%%%%%%%%%%%
\section{Numerical method and simulation setup}%
\label{sec:num-methods}
%%%%%%%%%%%%%%%%%%%%%%%%%%%%%%%%%%%%%%%%%%%%%%%%

\subsection{Numerical Code}
A detailed description of the simulation setup, the numerical methods and the physical processes covered in this study are given in \citet{SILCC1}. We therefore only provide a brief description here. The simulations are carried out with the astrophysical code \textsc{FLASH} in version~4 \citep{FLASH00,DubeyEtAl2008,DubeyEtAl2013}, which is an adaptive mesh refinement (AMR) code developed at the University of Chicago. We solve the magneto-hydrodynamic (MHD) equations using the five-wave Bouchut MHD solver HLL5R \citep{Bouchut2007, Bouchut2010, Waagan2009, Waagan2011}. We numerically solve the following set of equations:
\begin{align}
\frac{\partial\rho}{\partial t} + \nabla \cdot (\rho \mathbf{v}) &= 0\\
\frac{\partial(\rho\mathbf{v})}{\partial t} + \nabla\cdot\left[\rho\mathbf{v}\mathbf{v}^\mathrm{T} + \left(P + \frac{\mathbf{B}^2}{8\pi}\right)\mathbf{I} - \frac{\mathbf{B}\mathbf{B}^\mathrm{T}}{4\pi}\right] &= \rho \mathbf{g} + \dot{\mathbf{q}}_\mathrm{inj}\\
\frac{\partial E}{\partial t} + \nabla\cdot\left[\left(E + \frac{\mathbf{B}^2}{8\pi} + \frac{P}{\rho}\right)\mathbf{v} - \frac{\left(\mathbf{B}\cdot\mathbf{v}\right)\mathbf{B}}{4\pi}\right] &=\notag\\
\rho\mathbf{v}\cdot\mathbf{g} + \dot{u}_\mathrm{chem} + \dot{u}_\mathrm{inj}&\\
\frac{\partial \mathbf{B}}{\partial t} -\nabla\times\left(\mathbf{v}\times\mathbf{B}\right) = 0.
\end{align}
The total energy density is given by
\begin{equation}
  E = u + \frac{\rho\mathbf{v}^2}{2} + \frac{\mathbf{B}^2}{8\pi}.
\end{equation}
Here, $\rho$ is the mass density, $\mathbf{v}$ the velocity, $P=(\gamma-1)u$ the thermal pressure with $u$ being the internal energy density and $\gamma$ the adiabatic index of the gas, which we set to $5/3$. We do not account for any softening of the equation of state in molecular-dominated regions due to the influence of the rotational and vibrational degrees of freedom of the hydrogen molecule because this effect becomes important only when the temperature is high enough to populate the excited states. This requires a temperature of at least $100-200\,\mathrm{K}$, but almost all of the molecular gas we find in our simulations is cooler than this \citep{SILCC1}. The magnetic field strength is denoted with $\mathbf{B}$. The gravitational acceleration $\mathbf{g}$ combines self-gravity and the effects of an external potential as described below. The total energy density $E$ is coupled to the change in internal energy density brought about by radiative and chemical heating and cooling, $\dot{u}_\mathrm{chem}$. Finally, $\dot{\mathbf{q}}_\mathrm{inj}$ and $\dot{u}_\mathrm{inj}$ describe the mechanical and thermal injection rates per unit volume from SNe.

\subsection{Gravity}
For the gravitational forces we take into account self-gravity and a background potential due to the stellar component of the disc, $\mathbf{g} = \mathbf{g}_\mathrm{sg} + \mathbf{g}_\mathrm{ext}$.
We include self-gravity of the gas by solving the Poisson equation,
\begin{equation}
  \Delta\Phi = 4\pi G \rho,
\end{equation}
with a tree based method \citepWuenschEtAl, where $\Phi$ is the gravitational potential, $G$ is Newton's constant and $\rho$ is the gas density. The external gravitational acceleration due to the stellar component in the galactic disc is modelled with an isothermal sheet originally proposed by \citet{Spitzer1942}, where the distribution function of stars is Maxwellian. For our setups we choose a stellar surface density of $\Sigma_*=30\,M_\odot\mathrm{pc}^{-2}$ comparable to the solar neighbourhood \citep{FlynnEtAl2006,BovyRixHogg2012} and a vertical scale height of $z_\mathrm{d}=100\,\mathrm{pc}$. We neglect any gravitational effects from dark matter because the expected local dark matter density ($\rho_\mathrm{DM}\sim10^{-3}\,M_\odot\,\mathrm{pc}^{-3}$) is more than an order of magnitude lower than our average density (gas plus stars) in a central volume of $(500\,\mathrm{pc})^3$, which is $\langle\rho_\mathrm{gas}+\rho_*\rangle\approx0.05\,M_\odot\,\mathrm{pc}^{-3}$. In addition, the characteristic scale height of the dark matter profile is of the order of $10\,\mathrm{kpc}$ \citep{CheminDeBlokMamon2011}. We also neglect variations of the stellar potential in the $x$ and $y$ direction in order not to impose an initial pattern or a characteristic length scale during the dynamical and chemical evolution of the gas in the disc.

\subsection{Chemistry and cooling}
\label{sec:chemistry-cooling}

We model the chemistry of the ISM using a simplified but fast chemical network based on \citet{GloverMacLow2007a,GloverMacLow2007b}, \citet{GloverEtAl2010}, \citet{NelsonLanger1997}, and \citet{GloverClark2012b}. This network is designed to follow the chemical abundances of six species: H$^{+}$, H, H$_{2}$, C$^{+}$, CO and free electrons; note that for simplicity we assume that the abundance of neutral atomic carbon (C) is negligible in comparison to that of C$^{+}$ or CO. 

The thermal evolution of the gas is modelled using the detailed atomic and molecular cooling function described in \citet{GloverEtAl2010}, \citet{GloverClark2012b} and \citet{SILCC1}. This cooling function makes use of the information on the chemical composition of the gas provided by our chemical network, and hence is significantly more accurate than the simple analytical functions used in a number of previous studies \citep{WalchEtAl2011,MicicEtAl2013}.

We assume a uniform interstellar radiation field (ISRF), which we scale linearly with the number of massive stars and thus the SN rate (see Sec.~\ref{sec:SN-energy-input}). For our fiducial models with a SN rate of $15\,\mathrm{Myr}^{-1}$ we use a value of $G_0=1.7$ \citep{Draine1978} in units of the Habing field, $G_0$ \citep{Habing1968}. The ISRF is attenuated in dense regions due to gas and dust shielding based on the column density using the TreeCol algorithm \citep{ClarkGloverKlessen2012}. Full details of the implementation of this chemical network in FLASH~4 and the way in which it is coupled to our TreeCol-based treatment of molecular self-shielding and dust shielding can be found in \citet{SILCC1} and \citetWuenschEtAl.

In the simulations presented in this paper, we assume solar metallicity, with fixed elemental abundances of carbon, oxygen and silicon given by $x_{\rm C} = 1.41 \times 10^{-4}$, $x_{\rm O} = 3.16 \times 10^{-4}$ and $x_{\rm Si} = 1.5 \times 10^{-5}$, respectively \citep{SembachEtAl2000}. Initially, we start with all of the carbon in the form of C$^{+}$ and all of the oxygen in the form of O. Silicon is present in the form of Si$^{+}$, and as it does not participate in the chemical network, is assumed to remain in this form throughout the simulation.

\subsection{SN energy input}
\label{sec:SN-energy-input}

Feedback is provided by SNe, which is injected as thermal energy or as momentum input depending on the resolution and the density of the injection region. If the radius at the end of the Sedov-Taylor expansion phase (i.e. the radius at which the SN expansion changes from the non-radiative expansion phase to the radiative snowplough phase) is resolved with at least 4~cells, we inject an energy of $10^{51}\,\mathrm{erg}$ per SN explosion as thermal energy and let the code convert thermal into kinetic energy self-consistently. If the density in the volume under consideration is too high to resolve the Sedov-Taylor expansion phase with $\rho_\mathrm{crit}=3\times10^{-24}\,\mathrm{g\,cm}^{-3}$ in our simulations, injecting thermal energy would lead to over-cooling because the resulting net temperature in the injection region would be below $\sim10^6\,\mathrm{K}$, where the cooling rates are very high. The SNe would then have a negligible dynamical and thermal effect \citep[see, e.g.][]{AnninosNorman1994,StinsonEtAl2006,CreaseyEtAl2011}. Instead, we inject the expected net momentum the gas would be exposed to if the substructures of the SN explosion were resolved. The physical details we employ are described in \citet{BlondinEtAl1998}. For a detailed investigation of the SN positioning and energy injection methods in a very similar chemical environment we refer to \citet{GattoEtAl2015}.

The SNe are injected at a constant rate from the very beginning of the simulation. From the Kennicutt-Schmidt relation \citep[hereafter simply called KS relation]{Schmidt1959,Kennicutt1998} we derive a star formation rate (SFR) for our simulation box, which in turn can be converted into an SN rate using a fixed initial stellar mass function. We assume the formation of one massive star per $100\,M_\odot$ of stars. This yields a SN rate of $15\,\mathrm{Myr}^{-1}$. We account for the variations and uncertainties in the KS relation \citep{BigielEtAl2008,SchrubaEtAl2011,LeroyEtAl2013,ShettyKellyBigiel2013,ShettyEtAl2014} by varying the SN rate by a factor of three above and below the standard value.

The positions of the SNe are chosen in different ways. We place them either in the density peaks (peak driving), at random positions in the $xy$-plane and a Gaussian distribution in $z$ with a standard deviation of $50\,\mathrm{pc}$ (random driving), or alternating between the two modes with half of the SNe exploding in density peaks and half at random positions (mixed driving). The random positions are determined beforehand and stored in a SN sequence to ensure the same positions in all simulations which use the random driving mode and the KS SN rate. In addition, we perform three runs with clustered SNe.

For the runs with clustered SNe we define different vertical scale heights for SNe of type~Ia and II, similar to \citet{deAvillezBreitschwerdt2004} or \citet{JoungMacLow2006}. We split the total number of SNe into a fraction of 20\% type~Ia SNe with a scale height of $325\,\mathrm{pc}$ and 80\% SNe of type~II with a scale height of $50\,\mathrm{pc}$ \citep{MillerScalo1979, Heiles1987, TammannLoefflerSchroeder1994}. Among the type~II SNe, $3/5$ explode in a clustered environment, the rest of the type~II SNe explode as individual SNe at random positions. For the clustered SNe we assume a total cluster lifetime of $40\,\mathrm{Myr}$, draw the number of SNe per cluster, $N$, from a power-law distribution $\propto N^{-2}$ \citep{KennicuttEdgarHodge1989,McKeeWilliams1997,ClarkeOey2002} with a minimum of $N=7$ and a maximum of $N=40$ SNe per cluster, and spread the $N$ SNe with equal temporal explosion intervals $dt=40\,\mathrm{Myr}/N$ over the cluster's lifetime \citep{JoungMacLow2006}. Each cluster has a fixed position over its lifetime. When the next SN in the stratified box is due we determine the kind of explosion (type~Ia, type~II in cluster, type~II individual) according to their fractional abundance, and if it is a cluster type~II SN we find the one of the predefined clusters, whose next SN is closest to the simulation time. After all SNe of a cluster exploded the cluster is removed from the list. New clusters with new random positions are created (or better activated) if the current SN that is due according to the KS rate does not fit in the explosion rate of one of the existing active clusters.

In addition we perform a run with the KS SN rate but with all SNe clustered as type~II with a scale height of $50\,\mathrm{pc}$ as used in the case of random driving. This run allows us to distinguish between effects that come from SN clustering and effects that are due to type~Ia SNe.

The justification of SNe at random positions -- and therefore most likely in low-density environments (see Sec.~\ref{sec:morphology}) -- is based on several aspects. First, there is a non-negligible fraction of runaway OB stars \citep[e.g.][]{deWitEtAl2005} that can travel a few hundred parsecs before they explode as a SN. For our simulation box with an area of $0.5\times 0.5\,\mathrm{kpc}^2$ in the $xy$-plane a random positioning seems reasonable \citep[see discussion in][]{LiEtAl2015}. In addition, recent work by \citet{BressertEtAl2012} and \citet{OeyEtAl2013} suggests that some of OB stars is born in the field. Another important aspect is the numerical resolution. A SN that explodes in a low-density environment in our simulations would need to travel at least a few cells away from the imaginary high-density birth place of the progenitor star. At a resolution of $4\,\mathrm{pc}$ this distance can easily be $\sim40\,\mathrm{pc}$. Higher resolution would result in more resolved structures including cavities and voids next to dense filaments and cores, which are smoothed to $4\,\mathrm{pc}$ in our setups. Statistically, the same star would therefore need to travel much less in order to reach low densities. Besides the problems of the travel distance of massive stars we also expect them to create low-density regions around them early on via protostellar outflows and later via radiation and stellar winds. This does not justify a random position, but it justifies the accompanying low-density environment (see also Gatto et al., in preparation).

\subsection{Initial conditions}
The initial density profile of the ISM in our model is uniform in the $xy$-plane and follows a Gaussian distribution in the vertical direction,
\begin{equation}
  \rho(z) = \rho_0 \exp\left\{-\left(\frac{z}{2z_0}\right)^2\right\},
\end{equation}
with $z_0=30\,\mathrm{pc}$ and $\rho_0=9\times10^{-24}\,\mathrm{g}\,\mathrm{cm}^{-3}$. We apply a density floor of $\rho_\mathrm{min}=10^{-28}\,\mathrm{g}\,\mathrm{cm}^{-3}$ to mimic the hot atmosphere around the disc. The resulting column density of $\Sigma=10\,M_\odot\,\mathrm{pc}^{-2}$ is similar to the Galactic value at the solar radius, $\Sigma\approx13\,M_\odot\mathrm{pc}^{-2}$ \citep{FlynnEtAl2006}.
The gas is initially at rest. We deliberately refrain from setting initial turbulent motions and density perturbations in order not to introduce specific scales or dynamical patterns. Instead, the SN driving together with the chemical evolution and the dynamics due to gravitational forces is allowed to form structures and motions self-consistently. The gas in the disc is initially atomic at a temperature of $4500\,\mathrm{K}$. The hot regions above and below the plane consist of ionized gas with a temperature of $T=4\times10^8\,\mathrm{K}$. We set the molecular gas fraction to zero everywhere in the box to follow the formation history of H$_2$. The disc is set up in thermal pressure equilibrium \citep{WangKlessenEtAl2010}. The presence of self-gravity and external potential thus result in a disc that is initially not in hydrostatic equilibrium. However, test simulations where we switch off cooling and chemical evolution show that gravitational effects only marginally change the density profile while reaching hydrostatic equilibrium. The effects of cooling are much more severe in driving the disc out of equilibrium.

In the magnetic run we assume that the field is initially oriented in the $x$ direction. The field strength scales with the density,
\begin{equation}
  B(z) = B_0\sqrt{\frac{\rho(z)}{\rho_0}},
\end{equation}
with the field strength at $z=0$ being $B_0=3\,\mu\mathrm{G}$. No tangled component is introduced.

\subsection{Simulation overview}

The overall numerical parameters are listed in Table~\ref{tab:sim-parameters}. Our simulated box spans $500\,\mathrm{pc}$ in $x$ and $y$ direction and extends to $\pm5\,\mathrm{kpc}$ in $z$ direction. The effective resolution is set to $\Delta x=4\,\mathrm{pc}$, corresponding to $128\times128$ cells in the $xy$-plane for $|z|<2\,\mathrm{kpc}$. For $|z|>2\,\mathrm{kpc}$ we reduce the resolution to $\Delta x= 8\,\mathrm{pc}$. Table~\ref{tab:simulation-overview} gives a list of all simulations discussed in the paper. The simulation name is composed of the surface density of the disc ($\Sigma=10\,M_\odot\,\mathrm{pc}^{-2}\rightarrow\mathrm{S10}$), the SN rate where \emph{lowSN}, \emph{KS} and \emph{highSN} refer to $1/3$, one and three times the KS SN rate. The SN driving mode is indicated by \emph{rand} (individual randomly placed SNe), \emph{peak} (individual SNe in density peaks), \emph{mix} (mixed mode with half the SNe in random places and half in density peaks) and \emph{clus} (clustered SNe with random positions of the clusters). In run \emph{clus2} all SNe are of type~II and clustered without single individual SN explosions. The suffix \emph{nsg} denotes a comparison run without self-gravity. The simulation including magnetic fields with an ordered field strength in $x$ direction of $3\,\mu\mathrm{G}$ in the midplane has \emph{mag} attached to the simulation name.

%%%%%%%%%%%%%%%%%%%%%%%%%%%%%%%%%%%%%%%%%
\begin{table}
  \caption{Simulation parameters}
  \begin{center}
    \begin{tabular}{lcl}
      Quantity & Symbol & Value\\
      \hline
      box size & $x\times y\times z$ & $0.5\times0.5\times\pm5\,\mathrm{kpc}$\\
      effective resolution & $\Delta x$ & $4\,\mathrm{pc}$\\
      gas surface density & $\Sigma$ & $10\,M_\odot\,\mathrm{pc}^{-2}$\\
      gas disc std dev & $z_0$ & $30\,\mathrm{pc}$\\
      central gas density & $\rho_0$ & $9\times10^{-24}\,\mathrm{g}\,\mathrm{cm}^{-3}$\\
      stellar surface density & $\Sigma_*$ & $30\,M_\odot\,\mathrm{pc}^{-2}$\\
      stellar disc scale height & $z_d$ & $100\,\mathrm{pc}$\\
      std dev. SN type Ia & $z_\mathrm{SNIa}$ & $325\,\mathrm{pc}$\\
      std dev. SN type II & $z_\mathrm{SNII}$ & $50\,\mathrm{pc}$\\
    \end{tabular}
  \end{center}

  \label{tab:sim-parameters}
\end{table}
%%%%%%%%%%%%%%%%%%%%%%%%%%%%%%%%%%%%%%%%%

%%%%%%%%%%%%%%%%%%%%%%%%%%%%%%%%%%%%%%%%%
\begin{table*}
  \begin{center}
    \begin{minipage}{11cm}
      \caption{List of simulations}
      \begin{tabular}{rlcccccc}
        no. & simulation name & SN rate  & $f_{\rm rand} $ & $f_{\rm clus} $ & $f_{\rm type~II} $  & $B$ & self-\\
        &          & (Myr$^{-1}$)  & & & & ($\mu\mathrm{G}$) & gravity\\
        \hline
        1 & S10-lowSN-rand     &  5  &  1.0 & 0.00 & 1.0 & --   & yes\\
        2 & S10-lowSN-peak     &  5  &  0.0 & 0.00 & 1.0 & --   & yes\\
        3 & S10-lowSN-mix      &  5  &  0.5 & 0.00 & 1.0 & --   & yes\\
        \hline
        4 & S10-KS-rand-nsg  & 15  &  1.0 & 0.00 & 1.0 & --   & no \\
        5 & S10-KS-rand     & 15  &  1.0 & 0.00 & 1.0 & --   & yes\\
        6 & S10-KS-peak     & 15  &  0.0 & 0.00 & 1.0 & --   & yes\\
        7 & S10-KS-mix      & 15  &  0.5 & 0.00 & 1.0 & --   & yes\\
        8 & S10-KS-clus     & 15  &  --  & 0.48 & 0.8 & --   & yes\\
        9 & S10-KS-clus2    & 15  &  --  & 1.00 & 1.0 & --   & yes\\
        10 & S10-KS-clus-mag3 & 15  &  --  & 0.48 & 0.8 & 3    & yes\\
        \hline
        11 & S10-highSN-rand    & 45  &  1.0 & 0.00 & 1.0 & --   & yes\\
        12 & S10-highSN-peak    & 45  &  0.0 & 0.00 & 1.0 & --   & yes\\
        13 & S10-highSN-mix     & 45  &  0.5 & 0.00 & 1.0 & --   & yes\\
        \hline                                            
      \end{tabular}
  
      \medskip Column 3 gives the supernova rate per Myr. The SNe are distributed with a certain fraction of random locations, $f_{\rm rand}$, which is given in column 4: $f_{\rm rand} =1.0$ corresponds to purely random SN driving, whereas $f_{\rm rand}=0.0$ corresponds to pure peak SN driving. Column 5 and 6 give the fraction of clustered SNe and the fraction of SNe of type~II, respectively. Column 7 contains the magnetic field strength and column 8 indicates whether self-gravity is switched on or off. We use the same notation for the simulations as \citet{SILCC1}.
    \end{minipage}
  \end{center}
  \label{tab:simulation-overview}

\end{table*}
%%%%%%%%%%%%%%%%%%%%%%%%%%%%%%%%%%%%%%%%%%%%%%%%%%%%%%%%%%%%%%%%%%%%

\section{Morphological evolution}
\label{sec:morphology}

%%%%%%%%%%%%%%%%%%%%%%%%%%%%%%%%%%%%%%%%%%%%%%%%
\begin{figure*}
  \begin{minipage}{\textwidth}
    \begin{center}
      \includegraphics[width=\textwidth]{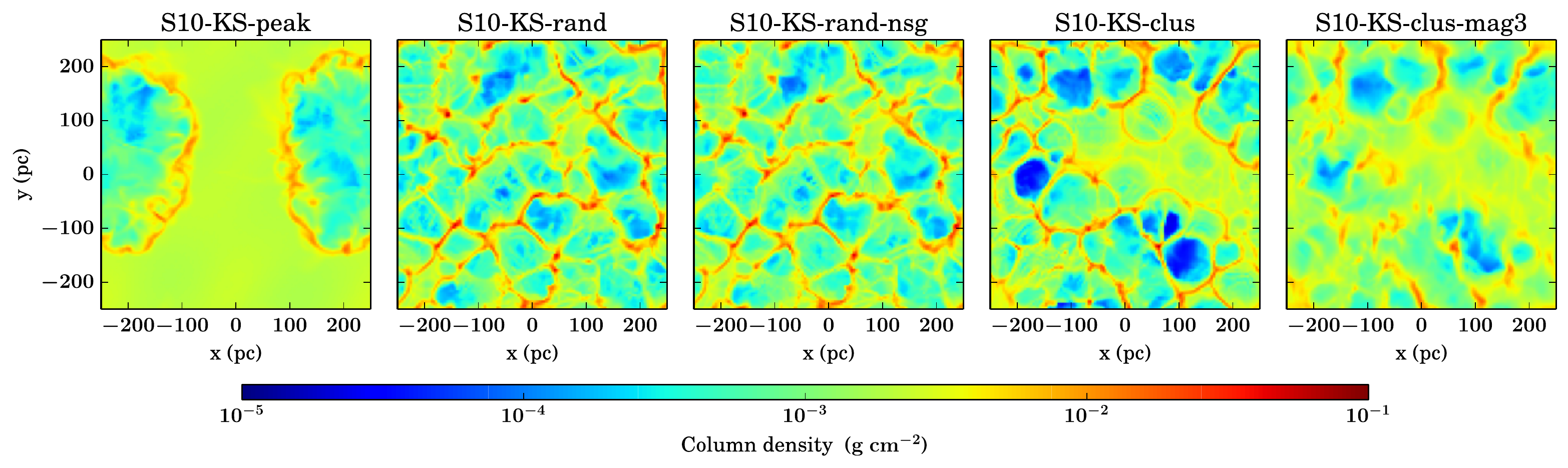}
    \end{center}
  \end{minipage}
  \caption{Early evolution of the column density for different SN driving modes at $t=10\,\mathrm{Myr}$. Shown are from left to right S10-KS-peak, S10-KS-rand, S10-KS-rand-nsg, S10-KS-clus and S10-KS-clus-mag3. The run with peak driving creates a superbubble due to the subsequent explosions of SNe in swept-up over-densities of the first SN. The forming structures in the clustered SN run are larger than in the case of individual random positions of the SNe because of multiple explosions at the same position. Self-gravity does not play a role at this stage of the simulation. The presence of magnetic fields causes dense structures to form later because of the stabilising magnetic pressure.}
  \label{fig:coldens-plane-early-evolution}
\end{figure*}
%%%%%%%%%%%%%%%%%%%%%%%%%%%%%%%%%%%%%%%%%%%%%%%%

%%%%%%%%%%%%%%%%%%%%%%%%%%%%%%%%%%%%%%%%%%%%%%%%
\begin{figure*}
  \begin{minipage}{\textwidth}
    \begin{center}
      \includegraphics[height=10.5cm]{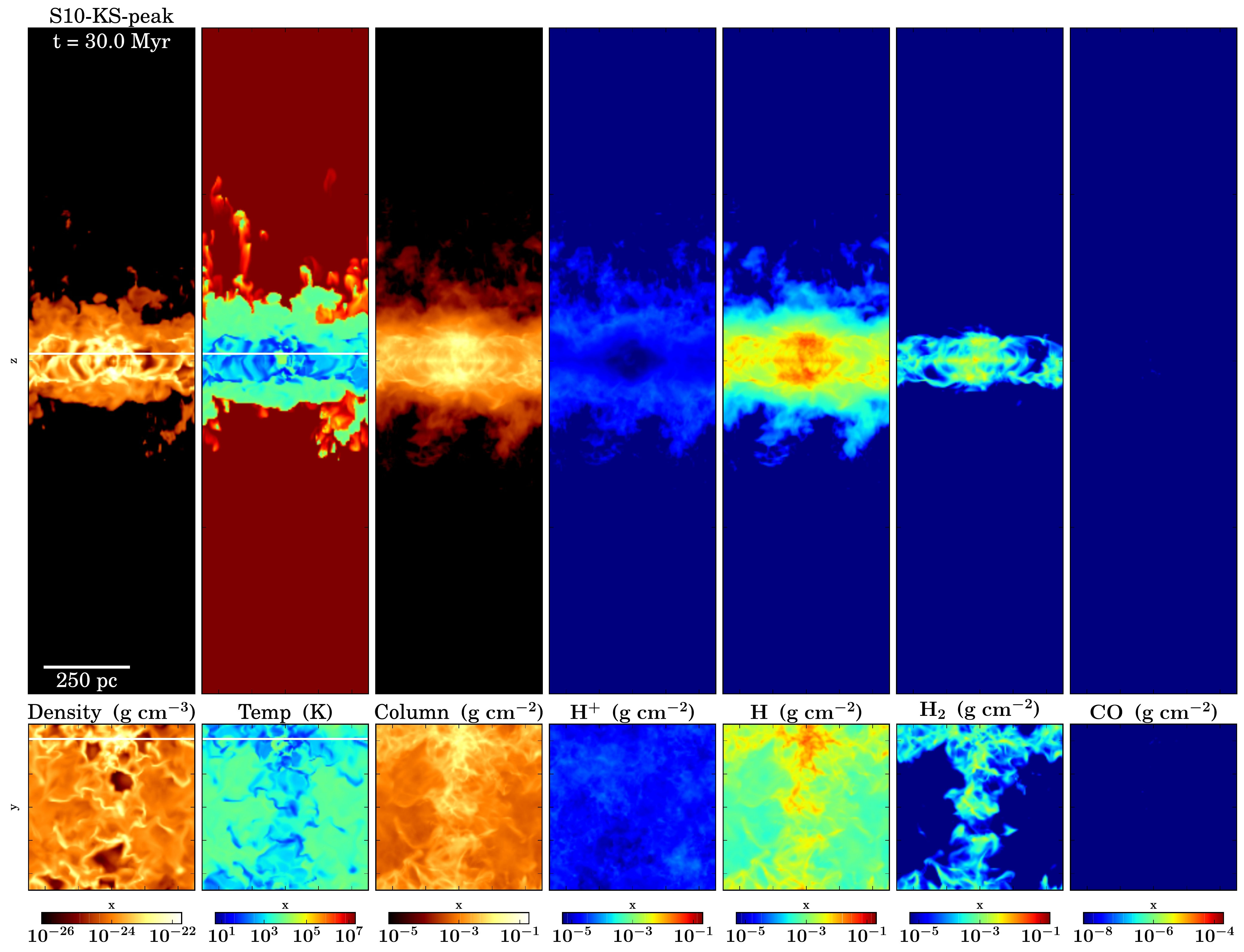}\\
      \includegraphics[height=10.5cm]{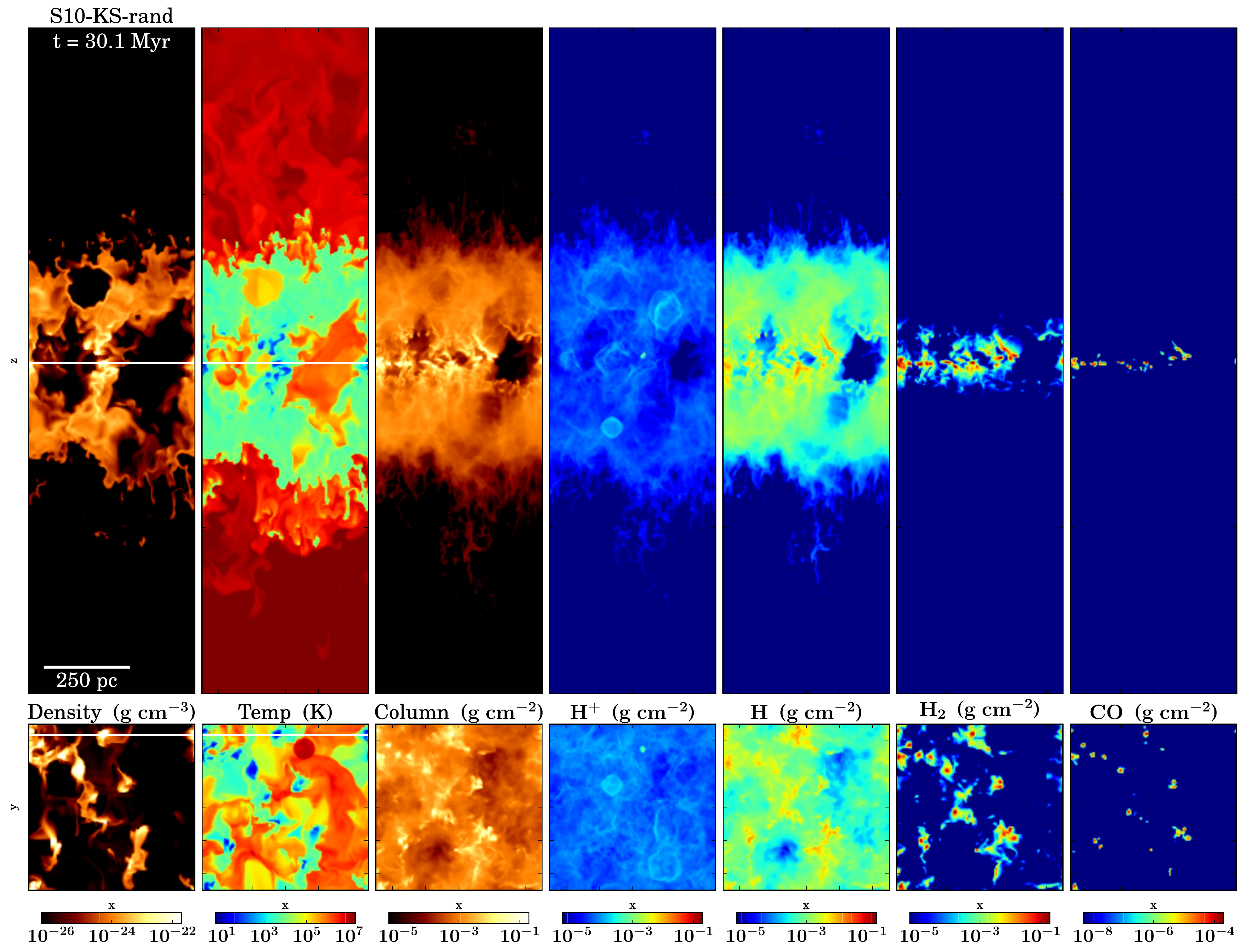}
    \end{center}
  \end{minipage}
  \caption{Overview of the simulations S10-KS-peak (top) and S10-KS-rand (bottom) at $t=30\,\mathrm{Myr}$. We plot (from left to right) a density and temperature cut through the position of the maximum density in the box followed by projections of the total gas density, H$^+$, H, H$_2$, and CO. The $y$ position of the cuts in the top panel is indicated by the white line in the bottom panel. Similarly, the $z$ position at which the cut in the lower panel is taken is indicated by the white line in the upper one. For peak driving (S10-KS-peak) the disc is very compact. The molecular component is dilute and most of the gas has intermediate and low temperatures ($\lesssim 10^4\,\mathrm{K}$). For random SNe (S10-KS-rand) the midplane contains more warm and hot gas. The density structures are clumpy and more molecular gas has formed.}
  \label{fig:coldens-S10-rand-peak-0300}
\end{figure*}
%%%%%%%%%%%%%%%%%%%%%%%%%%%%%%%%%%%%%%%%%%%%%%%%

%%%%%%%%%%%%%%%%%%%%%%%%%%%%%%%%%%%%%%%%%%%%%%%%
\begin{figure*}
  \begin{minipage}{\textwidth}
    \begin{center}
      \includegraphics[height=10.5cm]{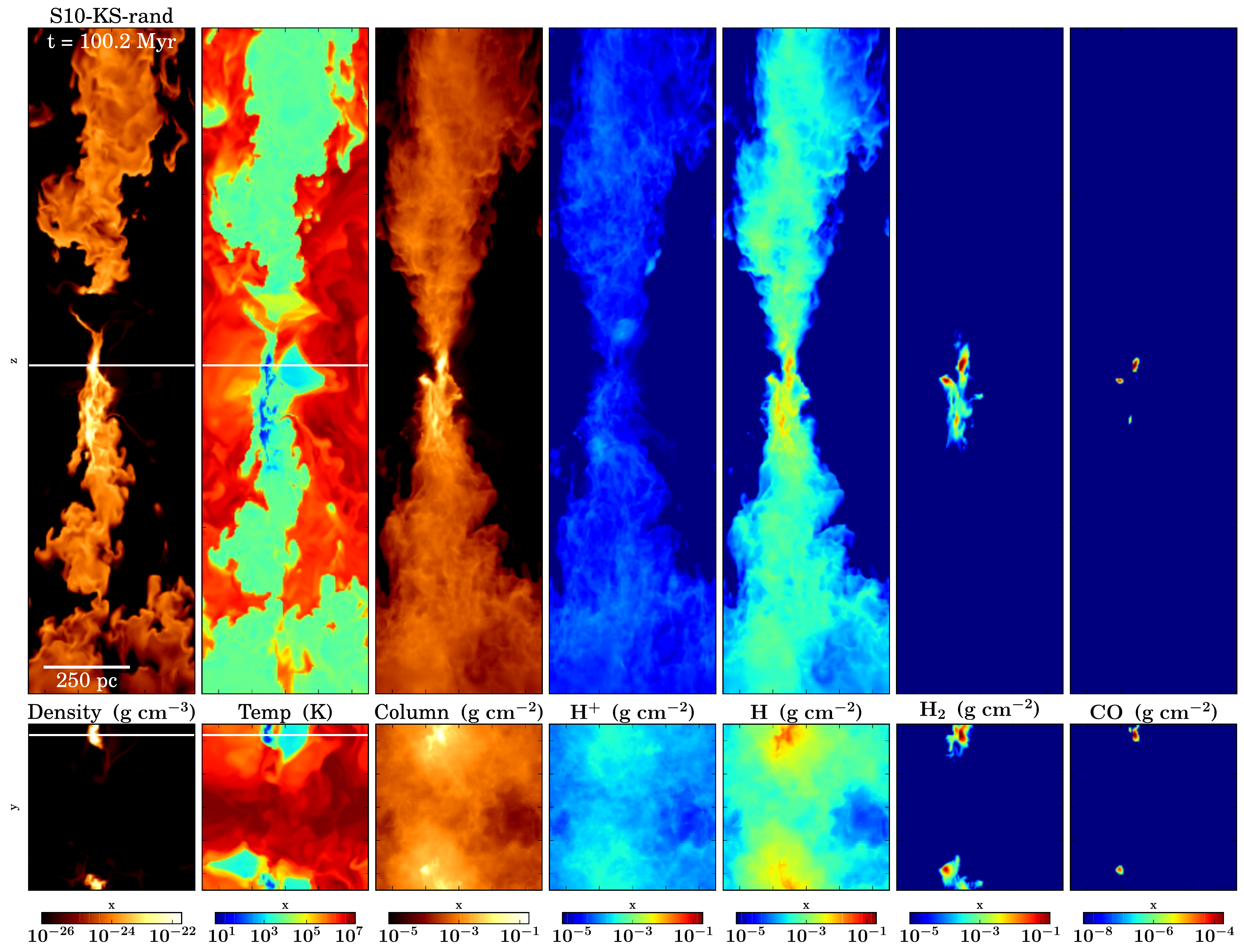}\\
      \includegraphics[height=10.5cm]{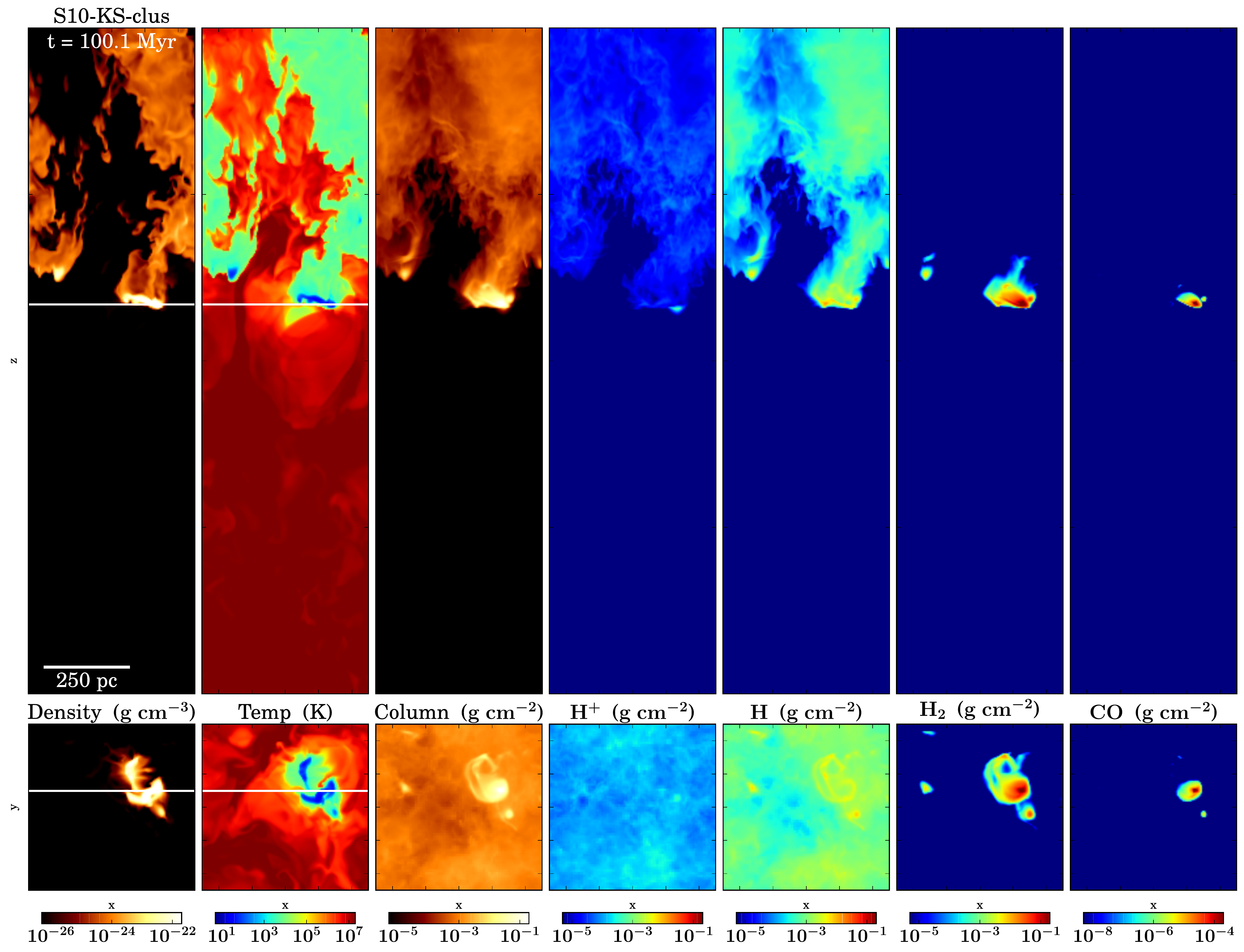}
    \end{center}
  \end{minipage}
  \caption{Overview of the simulations S10-KS-rand (top) and S10-KS-clus (bottom) at $t=100\,\mathrm{Myr}$. We plot (from left to right) a density and temperature cut through the position of the maximum density in the box followed by projections of the total gas density, H$^+$, H, H$_2$, and CO. The $y$ position of the cuts in the top panel is indicated by the white line in the bottom panel. Similarly, the $z$ position at which the cut in the lower panel is taken is indicated by the white line in the upper one. Initial filaments and small clouds merged to eventually form a few dense, massive GMCs. Simulation S10-KS-rand shows a dense column of mainly atomic gas above and below the plane. Temporal statistical asymmetries in the SN positions cause the formation of a GMC at $|z|\sim100\,\mathrm{pc}$ in simulation S10-KS-clus. Strong outflows that do not contain molecular gas have developed in both simulations.}
  \label{fig:coldens-S10-rand-clus-1000}
\end{figure*}
%%%%%%%%%%%%%%%%%%%%%%%%%%%%%%%%%%%%%%%%%%%%%%%%

%%%%%%%%%%%%%%%%%%%%%%%%%%%%%%%%%%%%%%%%%%%%%%%%
\begin{figure*}
  \begin{minipage}{\textwidth}
    \begin{center}
      \includegraphics[width=\textwidth]{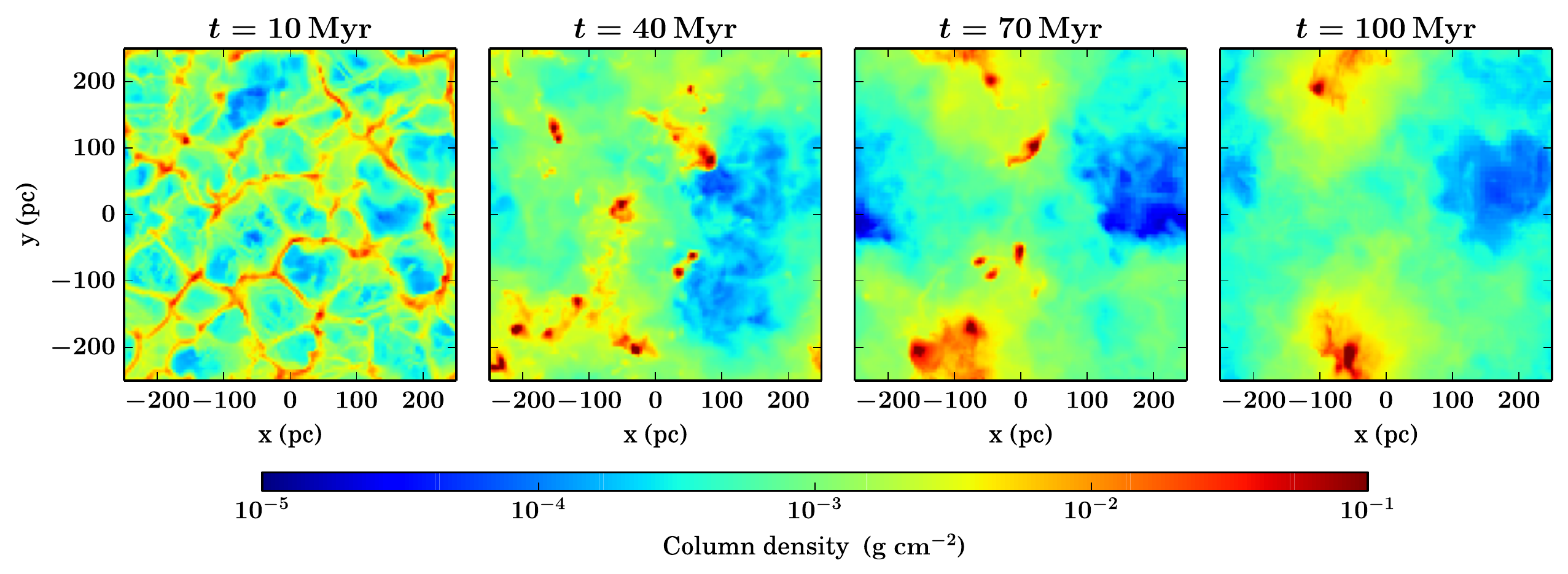}
    \end{center}
  \end{minipage}
  \caption{Evolution of the total gas column density at $t=10, 40, 70, 100\,\mathrm{Myr}$ for simulation S10-KS-rand. Over time the morphology changes from thin filaments and over multiple diffuse clouds that form at the intersections of filaments. The individual low-mass clouds both accumulate more gas and merge with other clumps to form a few very massive molecular clouds at the end of the simulation.}
  \label{fig:coldens-plane-structure-evolution}
\end{figure*}
%%%%%%%%%%%%%%%%%%%%%%%%%%%%%%%%%%%%%%%%%%%%%%%%

In Fig. \ref{fig:coldens-plane-early-evolution} we show a comparison of the face-on (integrated along the $z$ axis) column density after $10\,\mathrm{Myr}$ of evolution for runs with a KS star formation rate. At this early phase S10-KS-peak (left panel) shows clear signatures of the peak driving algorithm. The first SN creates an overdense shell where the subsequent SNe are placed automatically. This coherent clustered SN input sweeps up a lot of the gas and it takes about $\sim40\,\mathrm{Myr}$ to completely mix the gas in the disc. For random SN driving (S10-KS-rand, second panel in Fig. \ref{fig:coldens-plane-early-evolution}) the early morphology is very different. The gas develops a filamentary structure with column densities exceeding $0.1\,\mathrm{g\,cm}^{-2}$ (densities exceeding $10^{-21}\,\mathrm{g}\,\mathrm{cm}^{-3}$) in the filaments and dropping to values below $10^{-4}\,\mathrm{g\,cm}^{-2}$ ($10^{-27}\,\mathrm{g}\,\mathrm{cm}^{-3}$) in 'voids'. This early structure is created mainly by SN explosions. Self-gravity has little effect at this evolutionary stage (see S10-KS-rand-nsg). With clustered driving (S10-KS-clus) a similar filamentary morphology develops, however with larger structures caused by the larger number of spatially coherent SN explosions. The presence of the magnetic field delays the formation of dense structures. After $10\,\mathrm{Myr}$ the magnetic field run (S10-KS-clus-mag3, right panel in Fig. \ref{fig:coldens-plane-early-evolution}) shows less well developed structures at lower column densities.

In Fig.~\ref{fig:coldens-S10-rand-peak-0300} we show the edge-on and face-on morphology of the simulations after $t=30\,\mathrm{Myr}$ for simulations S10-KS-peak (top) and S10-KS-rand (bottom). We plot (from left to right) a density and temperature cut through the position of the maximum density in the box followed by projections of the total gas density, H$^+$, H, H$_2$, and CO. For peak driving (S10-KS-peak) the disc is more compact. The molecular component is dilute and most of the gas has intermediate and low temperatures ($\lesssim 10^4\,\mathrm{K}$). The accumulation of gas induced by the peak driving algorithm is clearly visible in the face-on plots. The structure of the disc does not change substantially during the evolution and no outflows are launched.

Already after $30\,\mathrm{Myr}$ the disc is significantly thicker (in particular for neutral and ionized hydrogen) with random supernova driving (S10-KS-rand) and the densities and temperatures indicate a noticeably larger volume filling fraction of hot ($\gtrsim 10^5\,\mathrm{K}$) gas near the disc plane \citep[for a quantitative analysis see][]{SILCC1}. The same qualitative behaviour can be seen for simulations with random driving and low and high star formation rates. The expanding neutral and ionized components indicate the onset of an ISM outflow as discussed in more detail in Sec.~\ref{sec:outflows}. The H$_2$ and CO columns show a number of dense clumps of molecular gas in the disc midplane which by this time already contributes $\sim 40$ per cent to the total gas content (only $\sim 7$ per cent for S10-KS-peak, see \citealp{SILCC1}).

The differences between random (S10-KS-rand) and clustered (S10-KS-clus) driving are highlighted in Fig.~\ref{fig:coldens-S10-rand-clus-1000} at a later stage of the simulation at $t=100\,\mathrm{Myr}$. Clustered SN explosions have mainly two effects. They drive gas flows to larger heights above the disc (described in detail in Sec.~\ref{sec:outflows}) and generate more hot gas with a larger volume filling factor. Local fluctuations in the gas distribution combined with the random -- but statistically centred around $z=0$ -- positioning of the clustered SNe allow for an asymmetric evolution with respect to the $z=0$ plane. Dense regions start to form above and below the plane. Over time the SNe, whose vertical distribution remains centred around $z=0$, can thus drive gas asymmetrically out of the plane, which explains the offset position of the massive cloud. In the subsequent analysis we thus define the midplane as the position of the densest gas for all simulations except S10-KS-peak and S10-KS-rand-nsg, whose densest peaks fluctuate perceptibly. In the disc region the gas is more concentrated in clumps with a larger fraction of H$_2$ in the run with clustered SNe compared to one with individual random SNe. At $t=100\,\mathrm{Myr}$ the small molecular clouds that have formed initially in the filamentary ISM have merged into a few very massive molecular complexes. A time sequence of this process is shown in Fig.~\ref{fig:coldens-plane-structure-evolution} for S10-KS-rand. After stable molecular structures have formed they are pushed towards each other in the presence of turbulent motions at scales of $\gtrsim10^2\,\mathrm{pc}$. At smaller distances they gravitationally attract each other and merge into larger complexes. The merging clouds are not disrupted because most of the random SNe and clusters are located within the volume-filling hot phase. The morphological evolution indicates that the simulations do not reach an equilibrium configuration in the morphology (see Sec.~\ref{sec:discussion} for a discussion).

%%%%%%%%%%%%%%%%%%%%%%%%%%%%%%%%%%%%%%%%%%%%%%%%%%%%%%%%%
\begin{figure}
  \centering
  \includegraphics[width=8cm]{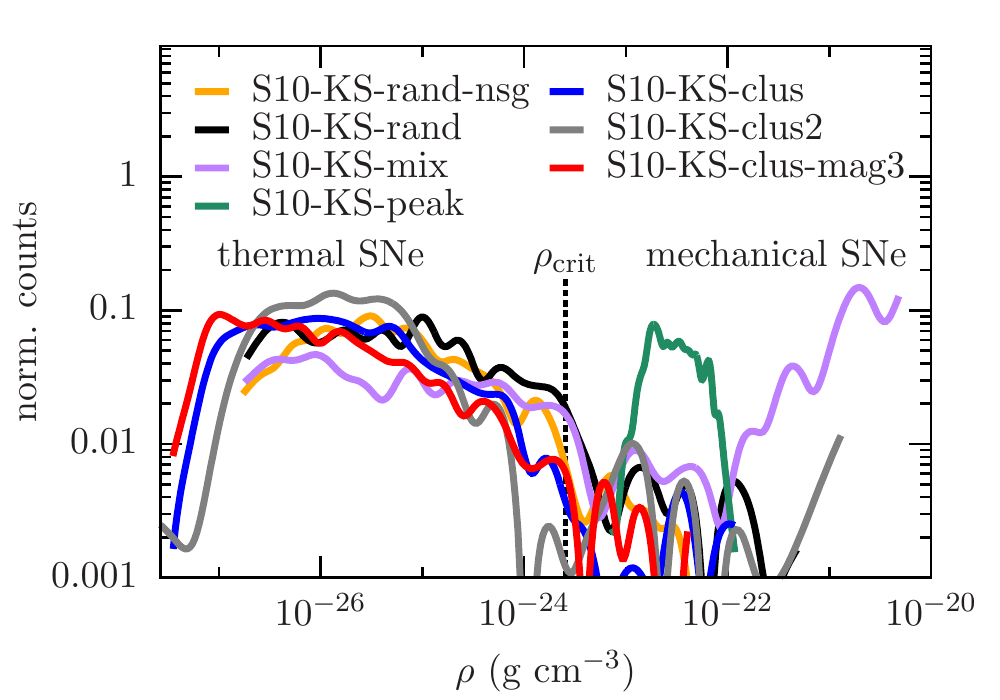}
  \caption{Distribution of the environmental density of the SNe for all simulations with KS SN rate averaged over the simulation time for $t\ge20\,\mathrm{Myr}$. Below the critical density, $\rho_\mathrm{crit}$, we inject thermal energy, while above it we use mechanical injection to avoid over-cooling. In run S10-KS-peak, all of the SNe explode in regions denser than $\rho_\mathrm{crit}$ with a peak of the distribution at $10^{-22}\,\mathrm{g}\,\mathrm{cm}^{-3}$, and so all of the SNe are injected as mechanical injection scheme. Random and clustered models are dominated by thermal explosions in the density range $10^{-27}-10^{-24}\,\mathrm{g}\,\mathrm{cm}^{-3}$.}
  \label{fig:SN-environment-density}
\end{figure}
%%%%%%%%%%%%%%%%%%%%%%%%%%%%%%%%%%%%%%%%%%%%%%%%%%%%%%%%%

The different morphological evolution of the ISM combined with the SN driving mode results in different environmental densities, in which the SNe explode. In Fig.~\ref{fig:SN-environment-density} we present the distribution of the environmental density, in which we inject the SNe. We average over time excluding the initial phase of the simulation ($t<20\,\mathrm{Myr}$). The vertical line indicates the threshold value for the thermal/kinetic injection mode. SNe at lower densities are injected as thermal energy. For higher densities we switch to a mechanical injection scheme (for details see Section~\ref{sec:SN-energy-input}). For peak driving all SNe turn out to be mechanical. Mixed SN positions show a broad bimodal distribution with environmental densities ranging from $10^{-26}\,\mathrm{g}\,\mathrm{cm}^{-3}$ to $10^{-21}\,\mathrm{g}\,\mathrm{cm}^{-3}$. The environmental densities are even higher than in the peak driving run because only half of the total number of SNe explodes in density peaks, which reduces the net counteracting force against gravitational attraction in dense regions. All random and clustered models are dominated by thermal energy injection in densities from $10^{-27}-10^{-24}\,\mathrm{g}\,\mathrm{cm}^{-3}$.

\section{Dynamical evolution}

%%%%%%%%%%%%%%%%%%%%%%%%%%%%%%%%%%%%%%%%%%%%%%%%%%%%%%%%%
\begin{figure*}
  \begin{minipage}{\textwidth}
  \centering
  \includegraphics[width=\textwidth]{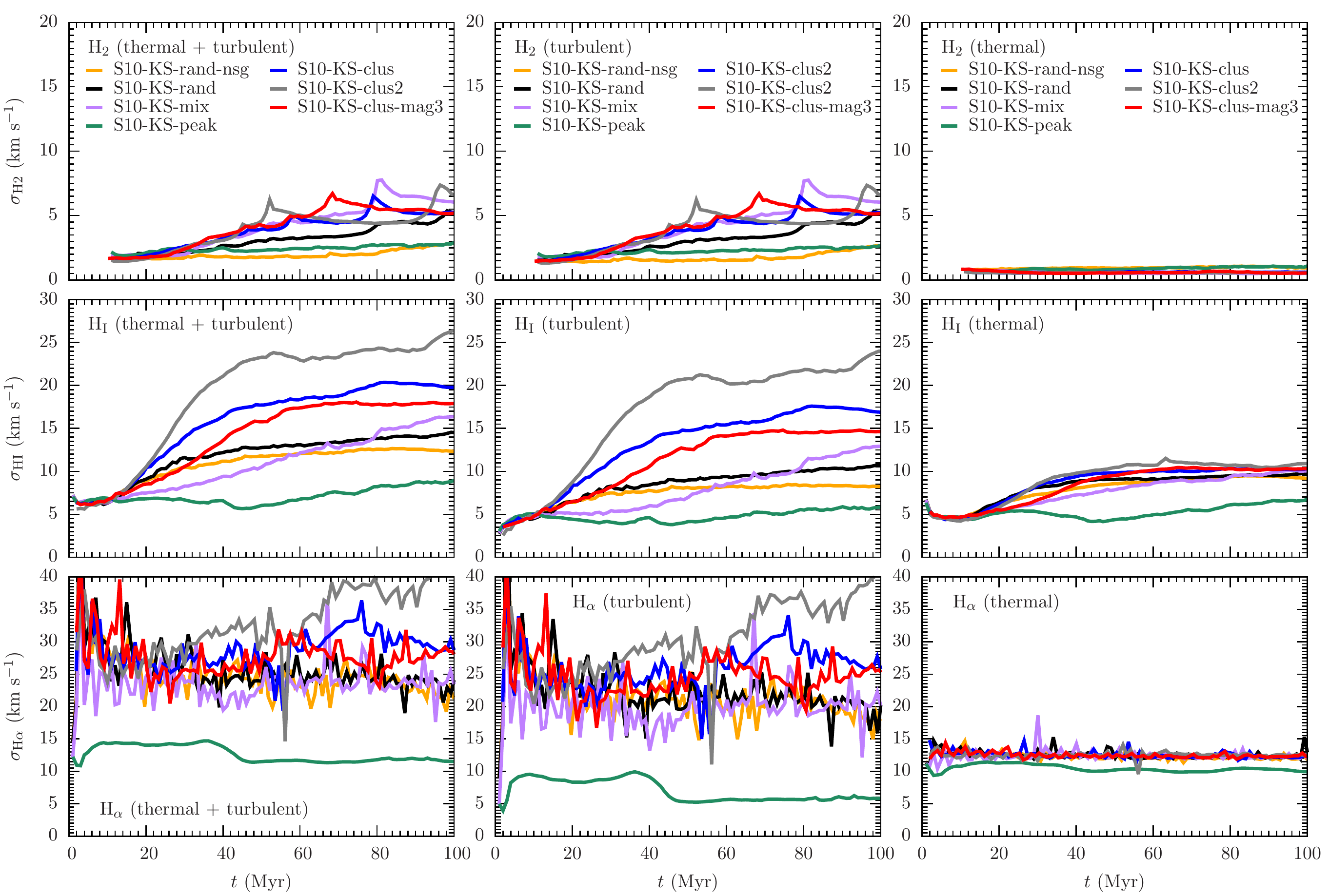}
  \caption{Velocity dispersions for all simulations with the KS SN rate. The left column depicts the total velocity dispersion (turbulent + thermal), the middle and the right column present the turbulent and thermal contribution. The top row shows mass-weighted values of $\sigma_\mathrm{H2}$, the middle and bottom rows show the observationally motivated estimates for $\sigma_\mathrm{HI}$ and $\sigma_{\mathrm{H}\alpha}$. The dynamics in the dense gas component, $\sigma_\mathrm{H2}$, is small for the simulation without self-gravity and for S10-KS-peak. The evolution of $\sigma_\mathrm{HI}$ and differs noticeably between random and clustered driving. For H$_2$ and H$_\alpha$ the turbulent component clearly dominates. In the case of H$_\textsc{i}$, the turbulent component only dominates for clustered driving. In all other runs both contributions $\sigma_\mathrm{turb}$ and $\sigma_\mathrm{therm}$ are similar.}
  \label{fig:S10-velocity-dispersion}
  \end{minipage}
\end{figure*}
%%%%%%%%%%%%%%%%%%%%%%%%%%%%%%%%%%%%%%%%%%%%%%%%%%%%%%%%%

To quantify the dynamical evolution of the gas we compute its one-dimensional velocity dispersion, $\sigma_s$, accounting for the thermal $\sigma_{s,\mathrm{therm}}$ and turbulent $\sigma_{s,\mathrm{turb}}$ contributions:   
\begin{equation}
  \label{eq:vel-disp-total}
  \sigma_{s} = (\sigma_{s,\mathrm{turb}}^2 + \sigma_{s,\mathrm{therm}}^2)^{1/2}.
\end{equation}
Here, $s$ indicates the ''gas phase'' under consideration. In our analysis, we consider two ways to partition the ISM into different phases. First, we use the traditional separation into four temperature regimes, ($T_1>3\times10^5\,\mathrm{K}$, $T_2\in\left[8000\,\mathrm{K};\,3\times10^5\,\mathrm{K}\right]$, $T_3\in\left[300\,\mathrm{K};\,8000\,\mathrm{K}\right]$, $T_4<300\,\mathrm{K}$), which are usually taken to represent the hot ionized medium (HIM), the warm neutral medium (WNM), thermally unstable gas between the WNM and the cold neutral medium (CNM), and finally the CNM itself, which is not separable from the even colder molecular gas in this approach. Our second method for identifying different ISM phases is more physically motivated and closer to observations. It is based on our ability to follow the chemical evolution of the gas (Sec.~\ref{sec:chemistry-cooling}). We estimate the flux in the H{\sc i} $21\,\mathrm{cm}$ line and the H$_\alpha$ line and use these as tracers of the atomic and ionized gas, respectively. Ideally, we would use a similar approach for the molecular component. However, the resolution in our current set of simulations is not sufficient to allow us to make accurate predictions for the CO emissivity, and we use the following approach for H$_{2}$ instead.

We calculate the mean turbulent velocity dispersion as 
\begin{equation}
  \label{eq:vel-disp}
  \sigma_{s,k,\mathrm{turb}} = \left(\frac{\sum_i(v_{k,i}-\overline{v}_k)^2\,m_{s,i}}{M_{s}}\right)^{1/2},
\end{equation}
where $k$ is the spatial direction ($x$, $y$, $z$) and $i$ is the index of the computational cell. The velocity $v_{k,i}$ is the velocity of cell $i$ and $\overline{v}_k=\left(\sum_im_iv_{k,i}^2\right)^{1/2}$ is the mean velocity in direction $k$. When computing the velocity dispersion of H$_{2}$, or of the different temperature components, we use simple mass weighting. In this case, $m_{s, i}$ represents the mass of the relevant phase in cell $i$, and $M_{s}$ represents the total mass in that phase in the entire computational volume. Alternatively, when computing the velocity dispersion of H{\sc i} or H$_{\alpha}$, we weight their contributions proportional to their estimated flux, as explained in more detail below. In this case, $m_{s, i}$ represents the flux from cell $i$ and $M_{s}$ the total flux. The mean turbulent one-dimensional velocity dispersion is then 
\begin{equation}
  \sigma_{s,\mathrm{turb}} = \left(\frac{1}{3}\sum_{k=1}^3\sigma^2_{s,k,\mathrm{turb}}\right)^{1/2}.
\end{equation}
The thermal contribution to the velocity dispersion (thermal broadening) is given by
\begin{equation}
  \sigma_{s,\mathrm{therm}} = \left(\frac{\sum_i v_{\mathrm{therm},i}^2\,m_{s,i}}{M_s}\right)^{1/2},
\end{equation}
with the thermal velocity $v_{\mathrm{therm},i} = (2k_\mathrm{B}T_i/\mu_im_\mathrm{H})^{1/2}$, the temperature, $T_i$, the mean molecular weight in cell $i$, $\mu_i$, and the mass of a hydrogen atom, $m_\mathrm{H}$.

The separation into temperature regimes is straightforward. To estimate the velocity dispersion for the $21\,\mathrm{cm}$ emission from atomic hydrogen, we assume the emission to be optically thin. This is a reasonable assumption for all but the highest column density lines-of-sight (\citealt{HeilesTroland2003,FukuiEtAl2014,MotteEtAl2014,BihrEtAl2015}) and allows us to determine $\sigma_{\rm HI,turb}$ directly from the atomic hydrogen number density provided by the chemical network. The H$_\alpha$ intensity is not directly proportional to the mass in H$^+$ as the emission decreases with temperature. To compute the H$_\alpha$ fluxes we follow the same procedure as outlined in \citet{GattoEtAl2015} accounting for collisional excitation and radiative recombination of ionized hydrogen \citep{DongDraine2011, Draine2011, KimEtAl2013}. We note that although most emission is expected from gas around $10^4$ K there can be contributions from radiative recombination at lower temperatures.

In Fig.~\ref{fig:S10-velocity-dispersion} we show the time evolution of the velocity dispersions $\sigma_\mathrm{H2}$, $\sigma_\mathrm{HI}$, and $\sigma_{\mathrm{H}\alpha}$ (from top to bottom) for the seven simulations with a KS star formation rate. The total values (turbulent plus thermal) are on the left, turbulent and thermal contributions (see Eq.~\ref{eq:vel-disp-total}) are given in the middle and right panels, respectively. We do not distinguish between intra- and inter-cloud dispersion. Molecular hydrogen (top panels of Fig.~\ref{fig:S10-velocity-dispersion}) starts forming from very cold gas only after $\sim 10\,\mathrm{Myr}$ (see \citealp{SILCC1}) and the turbulent component of the velocity dispersions slowly rise from $\sim 2$ to $\sim 6\,\mathrm{km}\,\mathrm{s}^{-1}$ for all runs except the two with peak driving (green line) and random driving without self-gravity (yellow line), whose values do not exceed $2.5\,\mathrm{km}\,\mathrm{s}^{-1}$ until the end of the simulation. The effect of self-gravity is therefore clearly visible in the molecular gas kinematics. As the H$_2$ gas is very cold ($T<300\,\mathrm{K}$), the total dispersion is dominated by turbulent motions with negligible contributions from thermal broadening.

The inferred velocity dispersions of neutral hydrogen (H$_\textsc{i}$, middle panels of Fig.~\ref{fig:S10-velocity-dispersion}) slowly rise from initial values of $\sim 5\,\mathrm{km}\,\mathrm{s}^{-1}$ to $\sim 10-25\,\mathrm{km}\,\mathrm{s}^{-1}$. Again, S10-KS-peak shows the smallest values for both the turbulent as well as the thermal components. The strong effect of self-gravity seen in the H$_2$ dynamics is almost absent (yellow line) as H$_\textsc{i}$ traces warm and more diffuse gas. For all runs with individual SN explosions (no clustering) the thermal broadening is comparable to the turbulent $\sigma$, whereas for clustered SNe the turbulent contribution clearly dominates. 

\begin{figure}
  \centering
  \includegraphics[width=8cm]{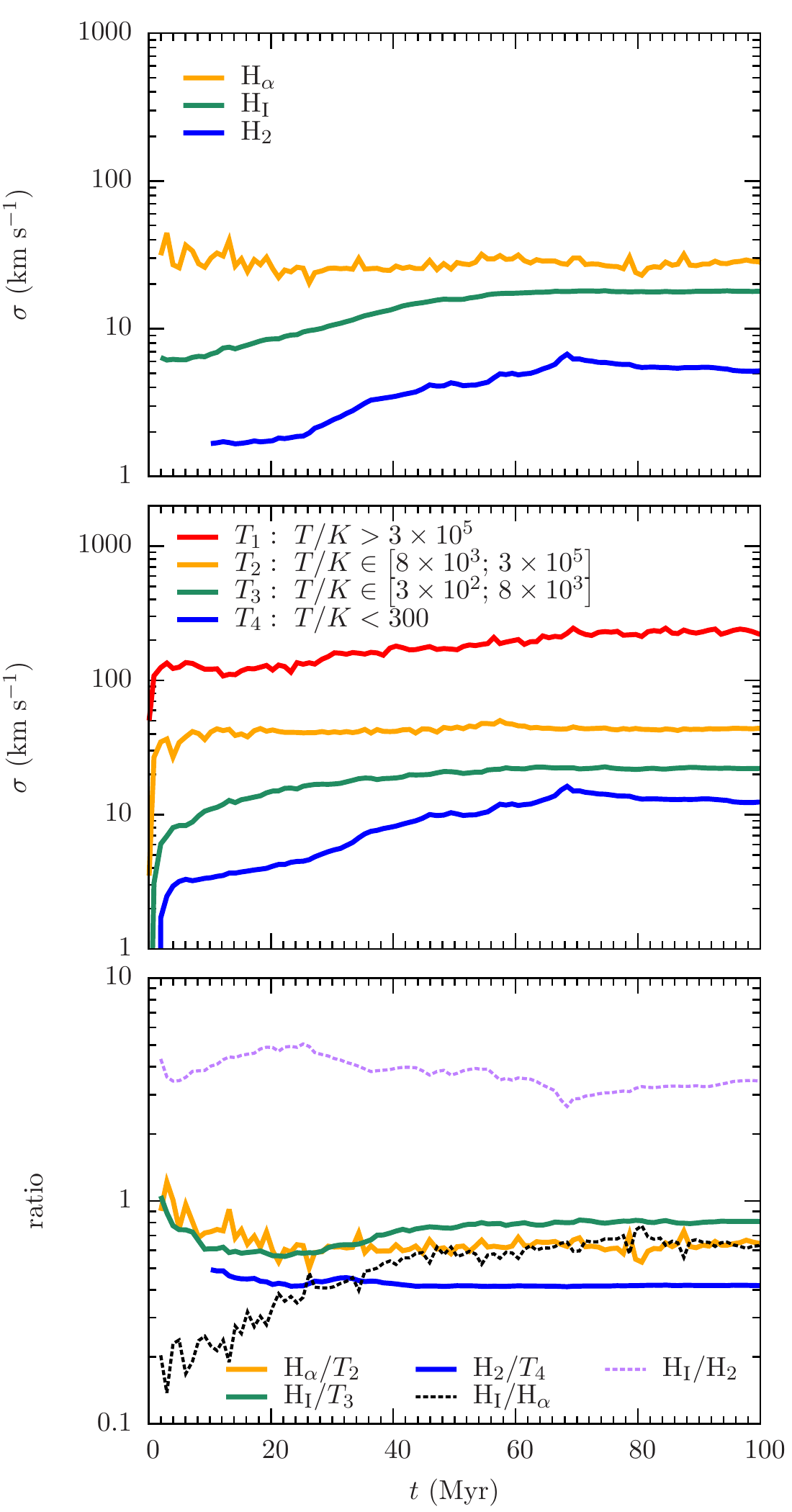}
  \caption{Time evolution of the velocity dispersion for simulation S10-KS-clus-mag3 for the chemical species and observational estimates (top), split up in temperature regimes (middle) and various ratios (bottom). The dynamics obtained from chemical species corresponds very well with the dynamics obtained from temperature cuts, indicated by the same colour in the top and middle panel. All ratios in the bottom panel show very little temporal variation.}
  \label{fig:S10-KS-clus-mag-sigma-temp-split}
\end{figure}

Observable H$_\alpha$ dispersions (bottom panels of Fig.~\ref{fig:S10-velocity-dispersion}) are roughly constant over time within the scatter with values of $\sim20-30\,\mathrm{km}\,\mathrm{s}^{-1}$. The mixed and random driving simulations show slightly lower values at the end ($\sim20-25\,\mathrm{km}\,\mathrm{s}^{-1}$), whereas the clustered driving runs reach $30-40\,\mathrm{km}\,\mathrm{s}^{-1}$ with the highest numbers for the simulation with only clustered SNe, S10-KS-clus2. A noticeable exception is again S10-KS-peak, in which basically all SNe are embedded in dense regions and little hot gas is accelerated. The fluctuations are stronger because for this hotter gas phase we start to see the direct impact of individual supernova explosions. For H$_\alpha$ dispersions the turbulent motions now dominate over the thermal broadening ($\sigma_{\mathrm{H\alpha,therm}} \simeq 12\,\mathrm{km}\,\mathrm{s}^{-1}$ for gas around $10^4\,\mathrm{K}$ as shown in the bottom panels of Fig.~\ref{fig:S10-velocity-dispersion}, see also Fig.~\ref{fig:S10-KS-clus-mag-sigma-temp-split}). Similar to H$_\textsc{i}$, there are no strong effects from self-gravity.  Magnetic fields (run S10-KS-clus-mag3) do not seem to perceptibly change the dynamical evolution of the systems, independent of the tracer.

Previous studies of the evolution of stratified discs did not include detailed models for the chemical evolution of the gas. They typically only focused on gas in different temperature regimes (e.g. \citealp{deAvillezBreitschwerdt2005}). We use what we consider our most realistic model, i.e. the run with magnetic field and clustered driving (S10-KS-clus-mag3), to investigate which temperature regime is probed best by the tracers we have just discussed. In the upper panel of Fig.~\ref{fig:S10-KS-clus-mag-sigma-temp-split} we collect the results from S10-KS-clus-mag3 (red lines in Fig.~\ref{fig:S10-velocity-dispersion}). In the middle panel we show the evolution of the gas dispersions in the five different temperature regimes. The hot ionized gas ($T_1 > 3\times10^5\,\mathrm{K}$) will be emitting in X-rays (not investigated here) and reaches a peak dispersion of $\gtrsim 200\,\mathrm{km}\,\mathrm{s}^{-1}$. The dynamics of the warm ionized gas ($\sigma_{\mathrm{T2}}$) is well traced by $\sigma_{\mathrm{H\alpha}}$. The same applies to the warm neutral gas in the temperature regime $T_3$ and the H$_\textsc{i}$ measurements (see the yellow (H$_\alpha/T_2$) and green (H$_\textsc{i}$/$T_3$) lines in the bottom panel of Fig.~\ref{fig:S10-KS-clus-mag-sigma-temp-split}). The velocity disperison of the cold gas ($T_4$) correlates well with the values for H$_2$, $\sigma_\mathrm{H2}/\sigma_{T4}=0.4$. Similarly, the ratio for H$_\textsc{i}/\mathrm{H_\alpha}$ shows remarkably small temporal scatter for $t>50\,\mathrm{Myr}$ with $\sigma_\mathrm{HI}/\sigma_{\mathrm{H}\alpha}=0.6$.

Our temperature-based velocity dispersions (middle panel of Fig.~\ref{fig:S10-KS-clus-mag-sigma-temp-split}) agree well with those reported by \citet{deAvillezBreitschwerdt2005}, \citet{JoungMacLowBryan2009}, \citet{KoyamaOstriker2009a}, \citet{KimKimOstriker2011}, or \citet{ShettyOstriker2012}, who all investigated the SN-driven ISM. The temperature-binned velocity dispersions found by \citet{PiontekOstriker2007} are slightly lower than ours. However, we note that these authors examined the magneto-rotational instability as a driver of turbulence rather than SNe.

%%%%%%%%%%%%%%%%%%%%%%%%%%%%%%%%%%%%%%%%%%%%%%%%
%%%%%%%%%%%%%%%%%%%%%%%%%%%%%%%%%%%%%%%%%%%%%%%%

Our simulations with random and clustered driving with self-gravity are also reasonable models for the observed ISM on spatial scales ranging between a few 10 to 100 parsecs. The dispersion in molecular hydrogen (directly determined from H$_2$ in the simulations) agrees well with observed (in CO) molecular gas complexes or giant molecular clouds (GMCs) in M31 ($\sigma=6.5\pm1.2\,\mathrm{km}\,\mathrm{s}^{-1}$, \citealp{ShethEtAl2008}) and GMCs in other star forming galaxies ($1-10\,\mathrm{km}\,\mathrm{s}^{-1}$, \citealp{BolattoEtAl2008}). More recently, \citet{DonovanMeyerEtAl2013} reported values of around $8\,\mathrm{km}\,\mathrm{s}^{-1}$ for spatially resolved observations of GMCs. To a similar level, our molecular dispersions agree with other GMC measurements in M31 \citep{Rosolowski2007}, the LMC \citep{FukuiEtAl2008}, and the SMC (\citealp{MizunoEtAl2001b}, see also the review by \citealp{FukuiKawamura2010}). \citet{IanjamasimananaEtAl2012} use data from the THINGS survey \citep{WalterEtAl2008} not only to estimate dispersions for the cold neutral medium ($\sigma=6.5\pm1.5\,\mathrm{km}\,\mathrm{s}^{-1}$) but also for the warm neutral medium. Their value of $\sigma=16.8\pm4.3\,\mathrm{km}\,\mathrm{s}^{-1}$ agrees well with our H$_\textsc{i}$ dispersions.

In our simulations $\sigma_\mathrm{HI}$ dominates over $\sigma_\mathrm{H2}$ throughout the simulation time with $\sigma_\mathrm{HI}/\sigma_\mathrm{H2}\approx3-4$. Assuming that $\sigma_\mathrm{CO}=\sigma_\mathrm{H2}$ this ratio is somewhat larger than observed values by \citet{CalduPrimoEtAl2013}, where most systems show $\sigma_\mathrm{HI}/\sigma_\mathrm{CO}\approx1-2$.

The estimates of our velocity dispersions for H$_\alpha$ are in agreement with observations. \citet{GreenEtAl2014} observe galaxies with resolved kinematics finding values in the range of $\sigma_\mathrm{H\alpha}\sim15-30\,\mathrm{km}\,\mathrm{s}^{-1}$ for Milky Way like systems.

%%%%%%%%%%%%%%%%%%%%%%%%%%%%%%%%%%%%%%%%%%
\section{Vertical structure and outflows}%
%%%%%%%%%%%%%%%%%%%%%%%%%%%%%%%%%%%%%%%%%%

\subsection{Vertical density structure}

%%%%%%%%%%%%%%%%%%%%%%%%%%%%%%%%%%%%%%%%%%%%%%%%
\begin{figure*}
  \begin{minipage}{\textwidth}
    \centering
    \includegraphics[width=8cm]{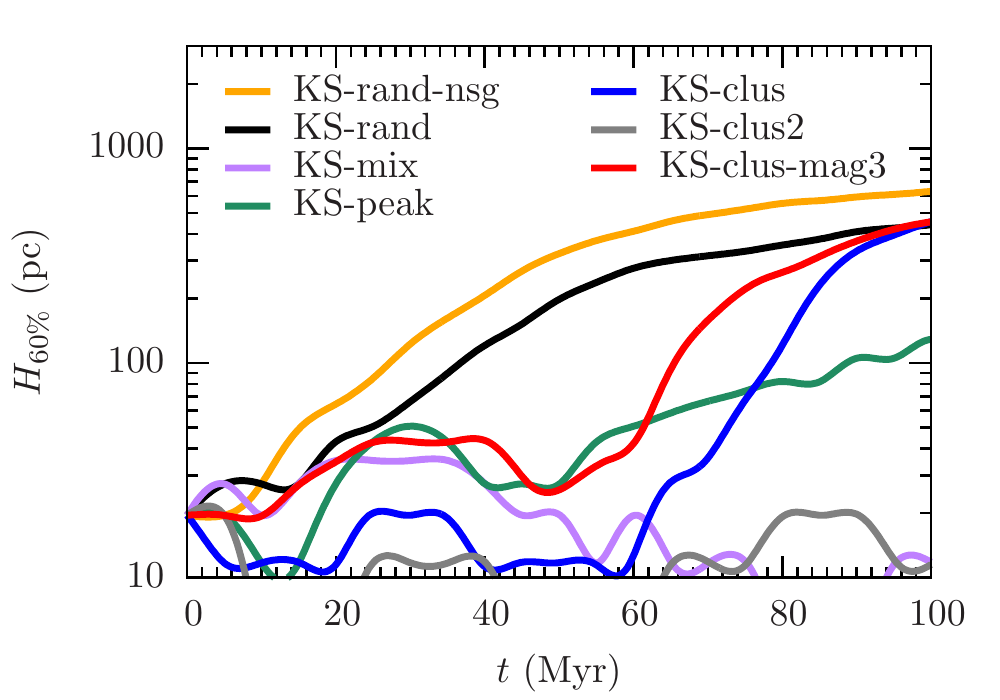}
    \includegraphics[width=8cm]{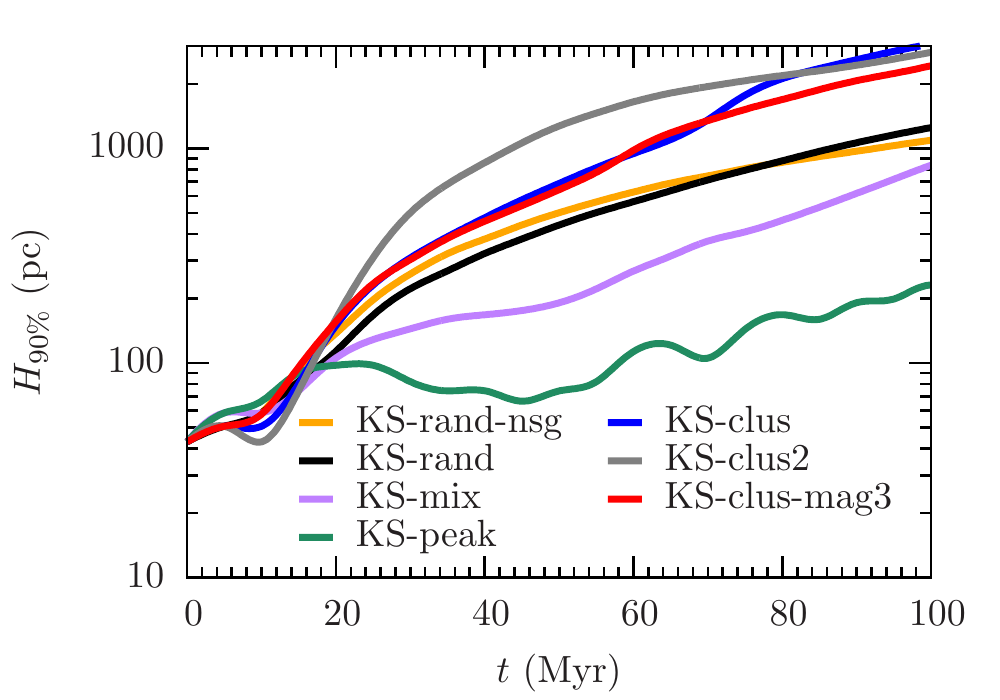}
  \end{minipage}
  \caption{Vertical heights of 60\% ($H_{60\%}$, left) and 90\% ($H_{90\%}$, right) enclosed mass over time. SN positions in density peaks result in compact discs with small values for $H_{90\%}$. Random and clustered SNe lead to a continuous expansion of the envelope ($H_{90\%}$) without any sign of a slowdown or turnover. Clustered driving models push the gas to noticeably higher altitudes. The height of the innermost 60\% of the gas differs between random and clustered driving modes. Without self-gravity the inner part of the disc starts expanding after only $10\,\mathrm{Myr}$. Self-gravity delays the expansion of the innermost 60\% of the mass. In the clustered runs the gas is pushed into dense structures more efficiently resulting in small values of $H_{60\%}$ until a significant fraction of the gas is expelled in outflows. For S10-KS-clus2 more than 60\% of the mass are confined in one GMC at the end of the simulation.}
  \label{fig:S10-vert-profiles-height}
\end{figure*}
%%%%%%%%%%%%%%%%%%%%%%%%%%%%%%%%%%%%%%%%%%%%%%%%

%%%%%%%%%%%%%%%%%%%%%%%%%%%%%%%%%%%%%%%%%%%%%%%%
\begin{figure*}
  \begin{minipage}{\textwidth}
    \centering
    \includegraphics[width=16cm]{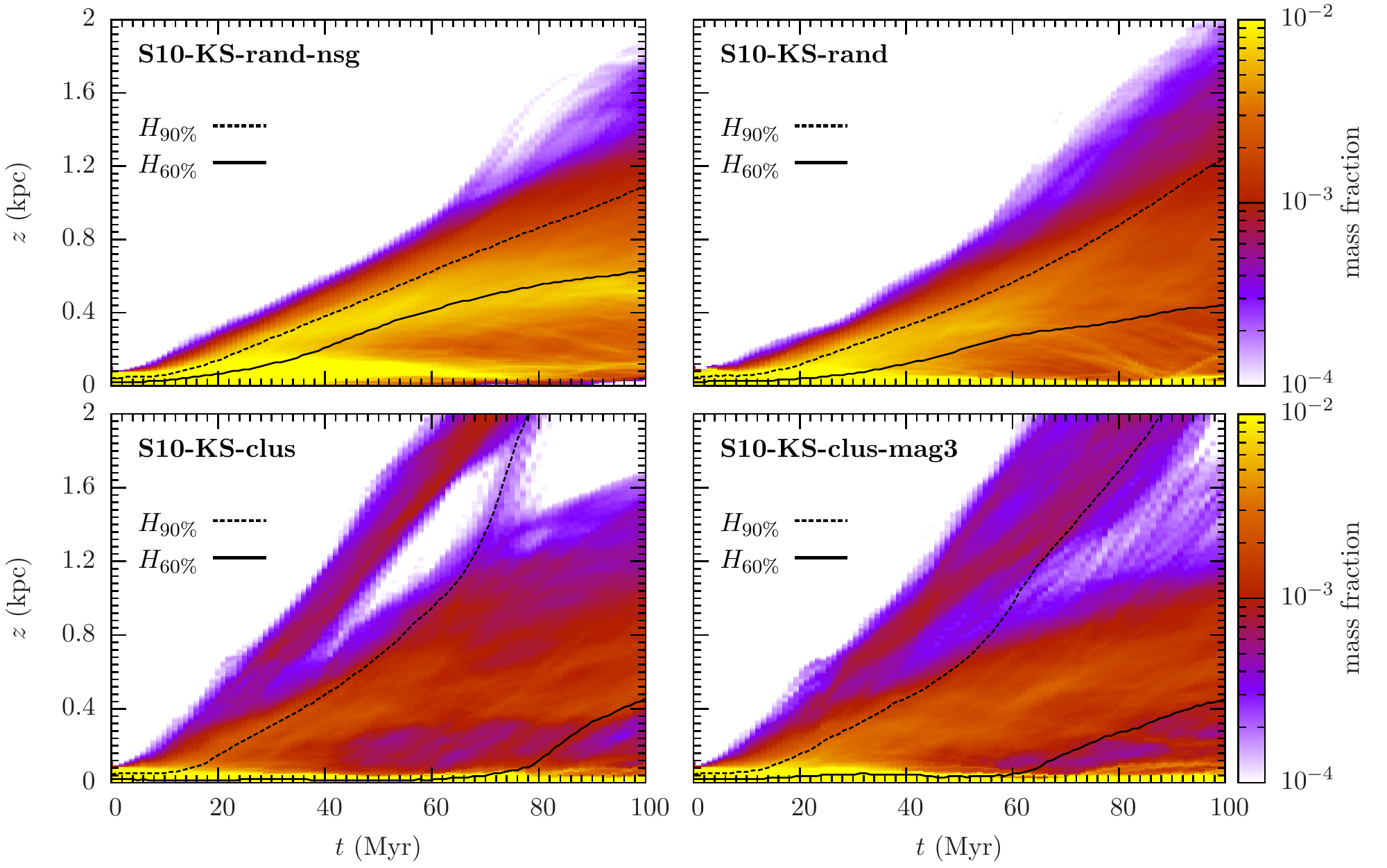}
  \end{minipage}
  \caption{Time evolution of the mass fraction as a function of height for simulations S10-KS-rand-nsg, S10-KS-rand, S10-KS-clus, and S10-KS-clus-mag3. Overplotted are the vertical heights of 60\% (solid line) and 90\% (dotted line) enclosed mass. The simulations with random SN positions have relatively massive and smooth vertical gas distributions, in particular at later times. The clustered SN models including 20\% of type~Ia SNe show low-density regions between the $H_{60\%}$ and $H_{90\%}$ line.}
  \label{fig:S10-mass-fraction-height}
\end{figure*}
%%%%%%%%%%%%%%%%%%%%%%%%%%%%%%%%%%%%%%%%%%%%%%%%

As the vertical density structure is complex, we do not attempt to fit it with a simple functional form, but instead determine the distance from the midplane enclosing 60\% ($H_{60\%}$) and 90\% ($H_{90\%}$) of the mass. For an exponential profile $\rho(z)=\rho_0\exp(-|z|/h)$ the total mass in the range $[-h, h]$ is about $60\%$, which motivates the choice of $H_{60\%}$. As almost all dense molecular gas is confined to the disc midplane the value of $H_{60\%}$ is connected to the dense structures and the molecular gas, whereas $H_{90\%}$ also includes the envelope of the disc.

Fig.~\ref{fig:S10-vert-profiles-height} shows the time evolution of $H_{60\%}$ (left) and $H_{90\%}$ (right) for all simulations with KS SN rate. In both runs with individual random SNe (S10-KS-rand and S10-KS-rand-nsg) the inner part of the disc ($H_{60\%}$) quickly expands. If self-gravity is switched off the expansion starts noticeably earlier and leads to a thicker disc over the entire simulation time. Clustered driving models can efficiently compress gas into dense filaments and clouds, which results in smaller values for $H_{60\%}$ for the first half of the simulation time. Runs S10-KS-clus and S10-KS-clus-mag3 eventually drive enough gas out of the midplane to reach $H_{60\%}\sim400\,\mathrm{pc}$, i.e. more than 40\% of the total mass is finally driven out of the disc. In the case of only clustered SNe (S10-KS-clus2) the efficient compression due to the coherent SNe results in more massive agglomerates of dense gas close to the midplane, which keeps the values for the 60\% mass limit below $\sim30\,\mathrm{pc}$ over the entire simulation time. Mixed SN driving ends up with similarly low numbers for $H_{60\%}$ but with a different evolution. The combination of half the SNe exploding in peaks and more volume filling random unclustered SNe first cause the disc to expand. The compact gas in the midplane only forms later when individual dense clouds merge and the fraction of SNe in density peaks cannot prevent the clouds from merging further. Simulation S10-KS-peak with a compact disc of mostly diffuse gas shows a very similar evolution for both $H_{60\%}$ and $H_{90\%}$.

In the evolution of the envelope of the disc ($H_{90\%}$) we notice a clear difference between individual random and clustered SNe driving. On the one hand clustered SNe result in more gas close to the midplane, on the other hand they can more efficiently drive an outflow, which causes the envelope to expand much faster. During the first $50\,\mathrm{Myr}$ the mixed driving model forms an extended diffuse disc, which does not allow the SNe do push gas to larger heights. In the second half of the simulation time outflows are eventually launched and the envelope can expand.

A more detailed view of the disc structure is shown in Fig.~\ref{fig:S10-mass-fraction-height}, where we plot the vertical distribution of the mass fractions as a function of time for simulations S10-KS-rand-nsg, S10-KS-rand, S10-KS-clus, and S10-KS-clus-mag3. We also overplot the heights for 60\% (solid line) and 90\% (dotted line) enclosed mass. The runs with random driving (upper panels) show smoothly distributed gas with no strong features in the profiles and an overall similar evolution of $H_{60\%}$ and $H_{90\%}$. The two runs with clustered driving (lower panels) show stronger density contrasts in the profile and a much larger difference between the evolution of the expanding envelope ($H_{90\%}$) and the extent of the dense part of the disc ($H_{60\%}$).

Comparing the vertical structure to other models is difficult because we cannot run the simulations for long enough to establish a full fountain cycle and a dynamical equilibrium on large scales over time. Outflows reshape the profiles until the end of the simulation. However, H$_2$ and CO form and remain close to the midplane of the disc in all our calculations, which allows us to compare those distributions to observations. We therefore compare H$_2$ scale heights with the heights of $60\%$ enclosed mass in H$_2$, $H_{\mathrm{H2},\,60\%}$, and similarly for $60\%$ of enclosed mass in CO, $H_{\mathrm{CO},\,60\%}$. In most simulations H$_2$ is very concentrated with $H_{\mathrm{H2},\,60\%}<30\,\mathrm{pc}$ except for simulations S10-KS-rand-nsg and S10-KS-peak for which $H_{\mathrm{H2},\,60\%}$ slowly increases to $\sim100\,\mathrm{pc}$ at $t=100\,\mathrm{Myr}$. However, for those two models the mass fraction of H$_2$ is below 5\% and the molecular gas is very diffuse. We find CO always embedded in the densest parts of the H$_2$ clouds and so $H_{\mathrm{CO},\,60\%}\lesssim H_{\mathrm{H2},\,60\%}$. Recent work by \citet{LangerPinedaVelusamy2014} finds a scale height based on an exponential fit of $47\,\mathrm{pc}$ for CO in the Milky Way. Observed scale heights in other galaxies can be as small as $40\,\mathrm{pc}$ for CO, but can also reach values of up to $200\,\mathrm{pc}$ \citep{CalduPrimoEtAl2013,YimEtAl2014}. A possible reason for why our molecular scale heights are on the low side of the observed range might simply be resolution. With our current resolution of $4\,\mathrm{pc}$, we are primarily sensitive to the largest GMCs, which predominantly form in the midplane, and we might miss the small, translucent high-latitude clouds that we see in the Milky Way. Indeed, observations of molecular clouds in the Milky Way show that the scale height of massive GMCs is considerably less than that of smaller molecular clouds \citep{StarkLee2005}, consistent with this explanation. Also, we would like to stress another limitation of our simulation setup. Our box represents only a small fraction of the galactic disc and is not exposed to large scale galactic dynamics. The gas in the box evolves in isolation and is not influenced by warps in the disc, inclined accretion flows, large scale fountain effects, and mergers. All of these effects are likely to act as an additional source of motion that might easily increase the vertical scale height of dense gas. In addition, other physical processes like CR pressure and radiative feedback are likely to enhance the dynamics of the disc and increase the scale height of the gas.

%%%%%%%%%%%%%%%%%%%%%%%%%%%%%%%%%%%%%%%%%%%%%%%%
\begin{figure*}
  \begin{minipage}{\textwidth}
    \centering
    \includegraphics[width=16cm]{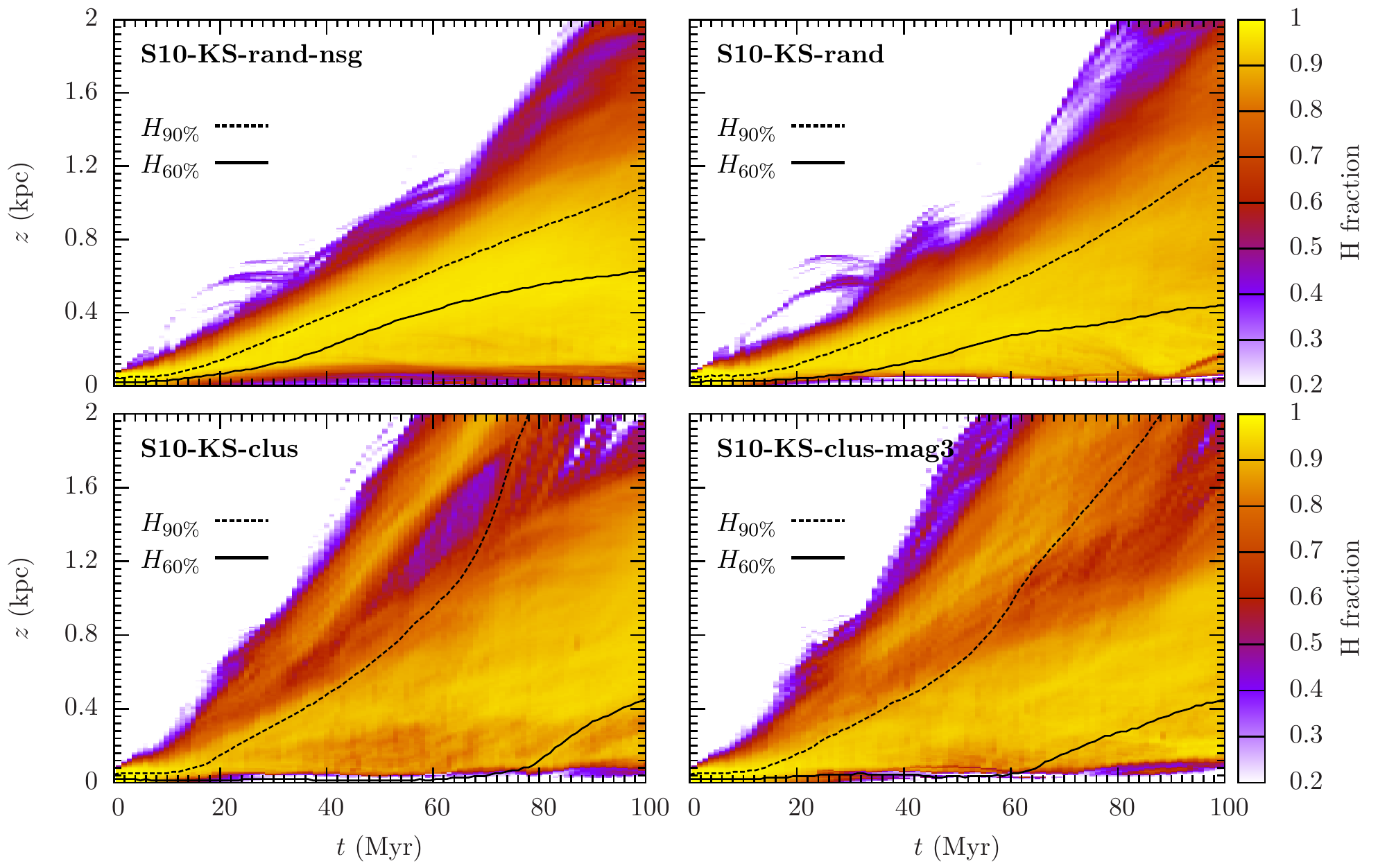}
  \end{minipage}
  \caption{H fraction over time for S10-KS-rand-nsg (top left), S10-KS-rand (top right), S10-KS-clus (bottom left) and S10-KS-clus-mag3 (bottom right). Overplotted are the heights of 60\% ($H_{60\%}$) and 90\% ($H_{90\%}$) enclosed mass. As H$_2$ only forms at very low heights, which can hardly be seen in this linear plot, the remaining mass of hydrogen is in the form of H$^+$. The outflows are overall dominated by atomic hydrogen. Individual random SNe lead to very high ratios of atomic to ionized hydrogen. In models with clustered SNe the composition locally drops to yield roughly equal proportions of H and H$^+$.}
  \label{fig:S10-vert-composition-height}
\end{figure*}
%%%%%%%%%%%%%%%%%%%%%%%%%%%%%%%%%%%%%%%%%%%%%%%%

The vertical distribution of the fraction of atomic hydrogen as a function of time is shown in Fig.~\ref{fig:S10-vert-composition-height} for simulations S10-KS-rand-nsg (top left), S10-KS-rand (top right), S10-KS-clus (bottom left) and S10-KS-clus (bottom right). The rest of the mass in the visible area is mainly H$^+$ because almost all the H$_2$ is confined within $|z|\sim100\,\mathrm{pc}$. Overplotted are the heights for $60\%$ ($H_{60\%}$) and $90\%$ ($H_{90\%}$) of the total enclosed gas mass. We notice a difference in the chemical structure between the random and clustered driving models. The former runs have smooth atomic hydrogen dominated gas distributions with a smoothly decreasing H fraction above the $H_{90\%}$ line. The latter simulations indicate a more complicated composition along the $z$ direction mainly due to the clustered SNe and the more powerful hot gas chimneys that are created in those simulations. The additional type~Ia SNe only show a minor effect, which we discuss in more detail in Section~\ref{sec:outflows}.

%%%%%%%%%%%%%%%%%%%%%%%%%%%%%%%%%%%%%%%%%%%%%%%%
\begin{figure}
  \centering
  \includegraphics[width=8cm]{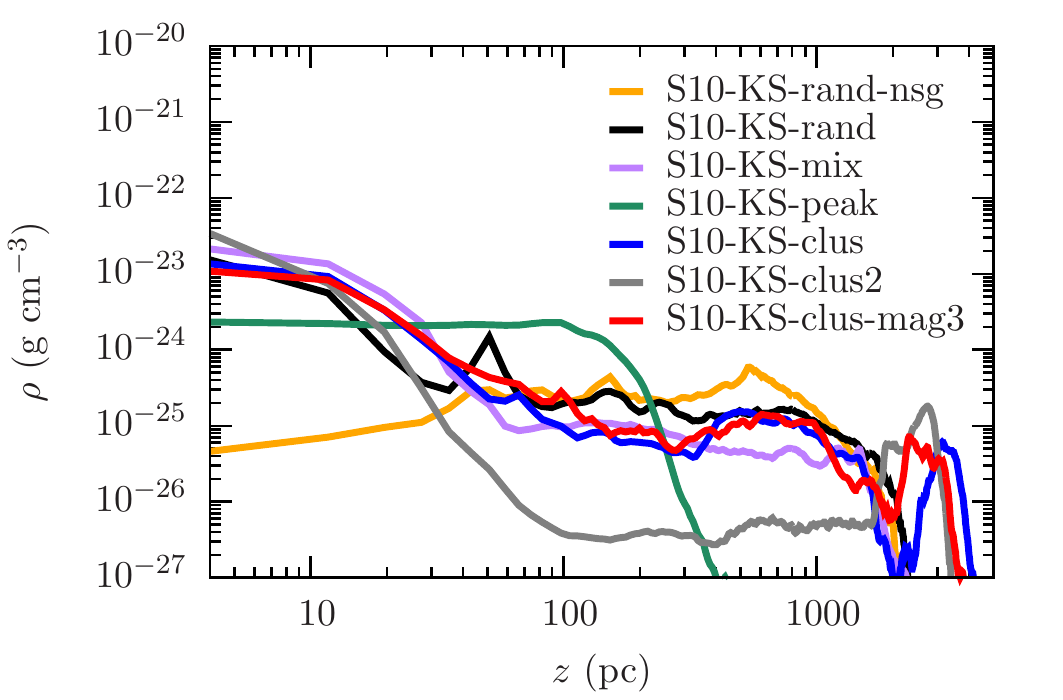}
  \caption{Vertical density profiles at $t=100\,\mathrm{Myr}$. The midplane region indicates very similar results with $\rho(z=0)\sim10^{-23}\,\mathrm{g}\,\mathrm{cm}^{-3}$ for all except two simulations. The one with peak SN driving shows a very compact smooth disc with a flat density profile up to $200\,\mathrm{pc}$. The central density is about an order of magnitude below the values of the other runs. The second noticeable exception is the random driving run without self-gravity with more than two orders of magnitude lower central densities. The efficiency of the SN clustering in shaping the environment is visible at heights above $z\sim100\,\mathrm{pc}$ where the density in S10-KS-clus2 is more than an order of magnitude lower compared to the runs with a lower clustering fraction.}
  \label{fig:S10-vert-profiles-dens}
\end{figure}
%%%%%%%%%%%%%%%%%%%%%%%%%%%%%%%%%%%%%%%%%%%%%%%%

Vertical density profiles averaged over $5\,\mathrm{Myr}$ of evolution are shown in Fig.~\ref{fig:S10-vert-profiles-dens} at $t=100\,\mathrm{Myr}$ (right). The central density of the simulation without self-gravity is more than two orders of magnitude smaller than in all runs including self-gravity. The peak driving model shows very flat density profiles up to $z\sim200\,\mathrm{pc}$, where the profile drops steeply. The simulations with only clustered SNe can efficiently drive outflows which is reflected in the lower densities above $z\sim100\,\mathrm{pc}$.

%%%%%%%%%%%%%%%%%%%%%%%%%%%%%%%
\subsection{Midplane pressure}%
%%%%%%%%%%%%%%%%%%%%%%%%%%%%%%%

%%%%%%%%%%%%%%%%%%%%%%%%%%%%%%%%%%%%%%%%%%%%%%%%
\begin{figure*}
  \begin{minipage}{\textwidth}
    \centering
    \includegraphics[width=16cm]{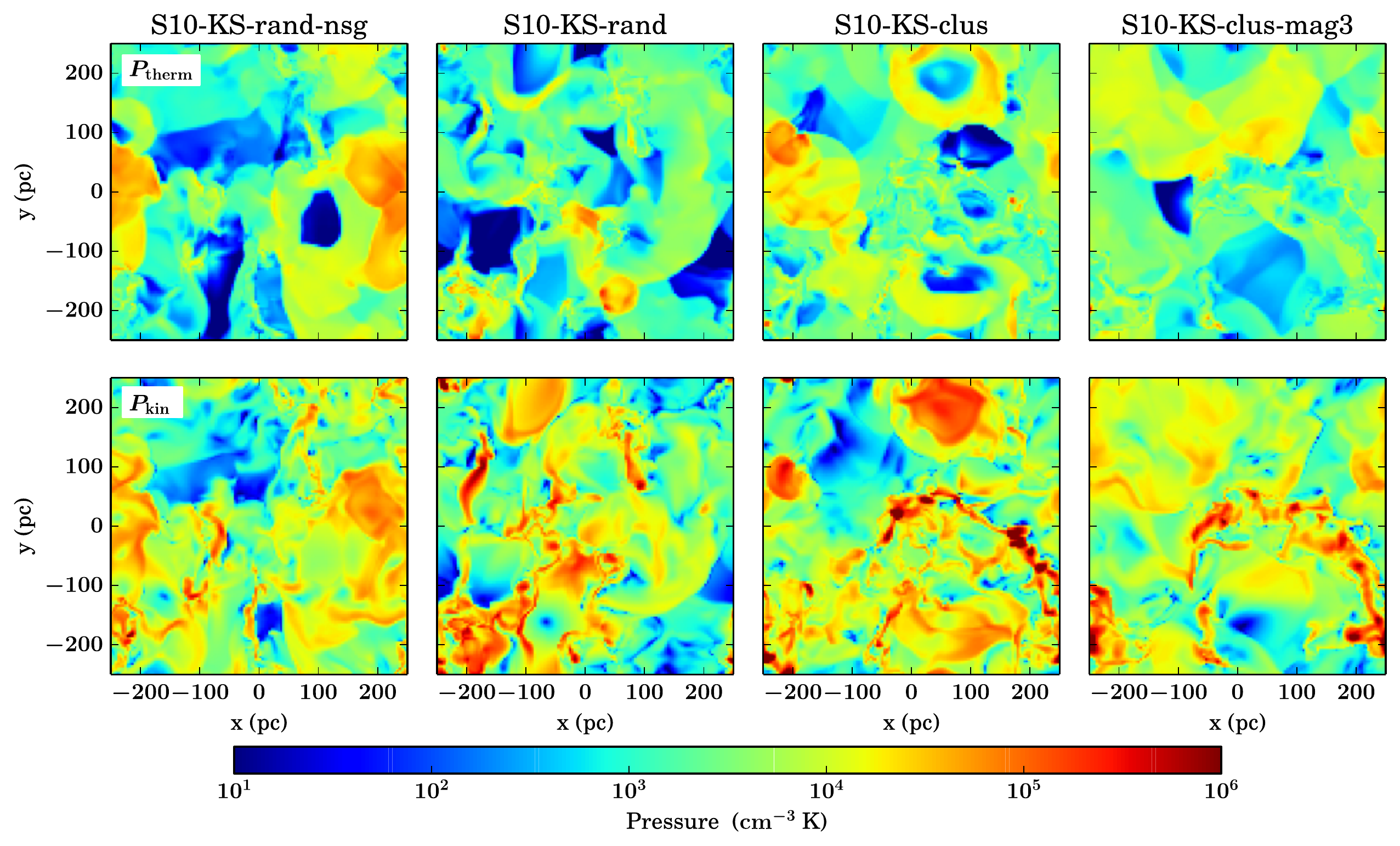}
    \includegraphics[width=16cm]{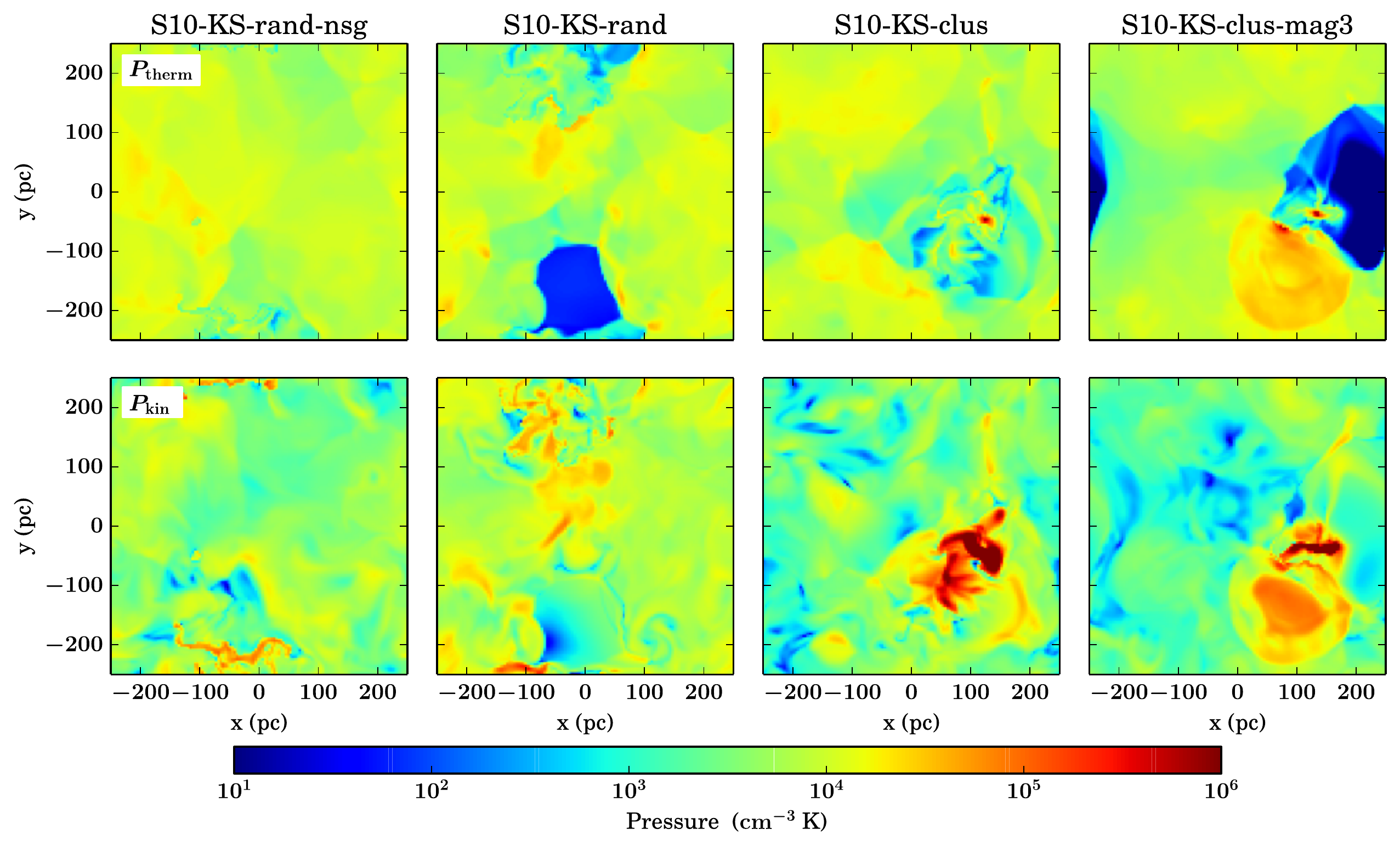}
  \end{minipage}
  \caption{Thermal and turbulent midplane pressures at $t=30\,\mathrm{Myr}$ (top) and $t=100\,\mathrm{Myr}$ (bottom). Overall the kinetic pressure dominates over the thermal one. At $t=30\,\mathrm{Myr}$ the pressure contrasts are large and reflect small-scale structures. Over time the small scale pattern in the pressure distribution evolves into large scale structures along with the coalescence of gas into fewer massive clouds towards the end of the simulation.}
  \label{fig:S10-profiles-pres-plane}
\end{figure*}
%%%%%%%%%%%%%%%%%%%%%%%%%%%%%%%%%%%%%%%%%%%%%%%%

%%%%%%%%%%%%%%%%%%%%%%%%%%%%%%%%%%%%%%%%%%%%%%%%
\begin{figure}
  \centering
  \includegraphics[width=8cm]{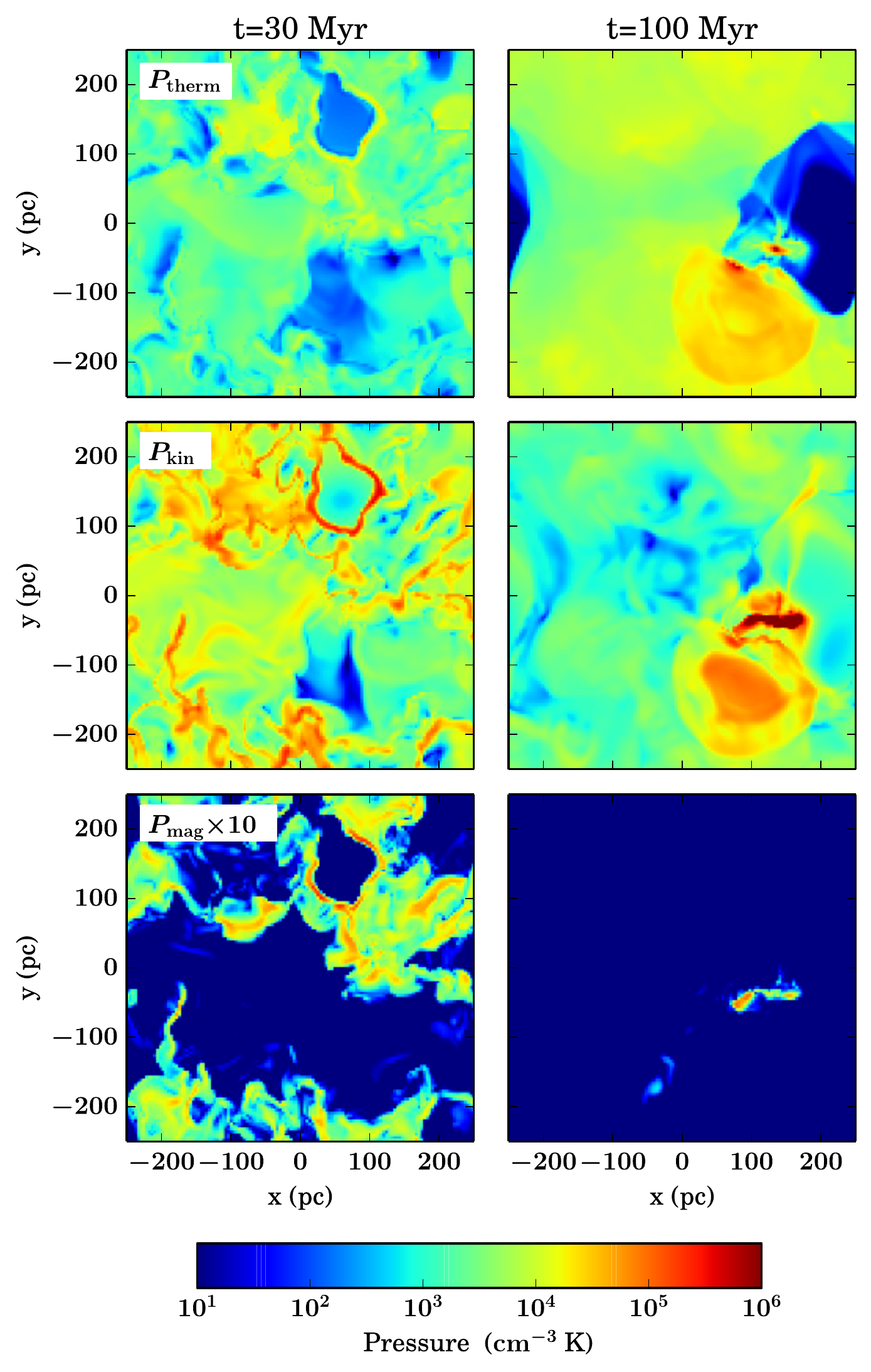}
  \caption{Midplane distribution of the thermal (top), kinetic (middle) and magnetic pressure (top) for simulation S10-KS-clus-mag3 at $t=30\,\mathrm{Myr}$ (left) and $t=100\,\mathrm{Myr}$ (right). The magnetic pressure is overall weaker and shown magnified by a factor of $10$. The ideal MHD approximation correlates the density structure with the field strength (flux freezing). The patches of high magnetic pressure thus trace the density field.}
  \label{fig:S10-profiles-mag-runs}
\end{figure}
%%%%%%%%%%%%%%%%%%%%%%%%%%%%%%%%%%%%%%%%%%%%%%%%

Gas motions are governed by thermal ($P_\mathrm{therm}$), kinetic ($P_\mathrm{kin}=\rho v^2$), and magnetic ($P_\mathrm{mag}=B^2/8\pi$) pressure, which are shown in cuts through the midplane for the non-magnetic runs in Fig.~\ref{fig:S10-profiles-pres-plane} at $t=30\,\mathrm{Myr}$ (top) and $t=100\,\mathrm{Myr}$ (bottom). At $t=30\,\mathrm{Myr}$ the values span five orders of magnitude, so the gas is clearly not in pressure equilibrium. For all simulations the kinetic pressure dominates over the thermal pressure, in particular in dense regions. In the subsequent evolution larger and larger structures form through merging of filaments and clouds, which is also visible in the pressure distribution at $t=100\,\mathrm{Myr}$.

In Fig.~\ref{fig:S10-profiles-mag-runs} we show a comparison of midplane-pressures for run S10-KS-clus-mag3. As $P_\mathrm{mag}$ is small compared to the other two components, we multiply the magnetic pressure by a factor of $10$ to better visualise the results. The flux-freezing of the ideal MHD approximation leads to a correlation of the magnetic pressure with the density. At $t=100\,\mathrm{Myr}$ the magnetic pressure contribution is confined to the dense structures in the midplane.

%%%%%%%%%%%%%%%%%%%%%%%%%%%%%%%%%%%%%%%%%%%%%%%%
\begin{figure*}
  \begin{minipage}{\textwidth}
    \centering
    \includegraphics[width=\textwidth]{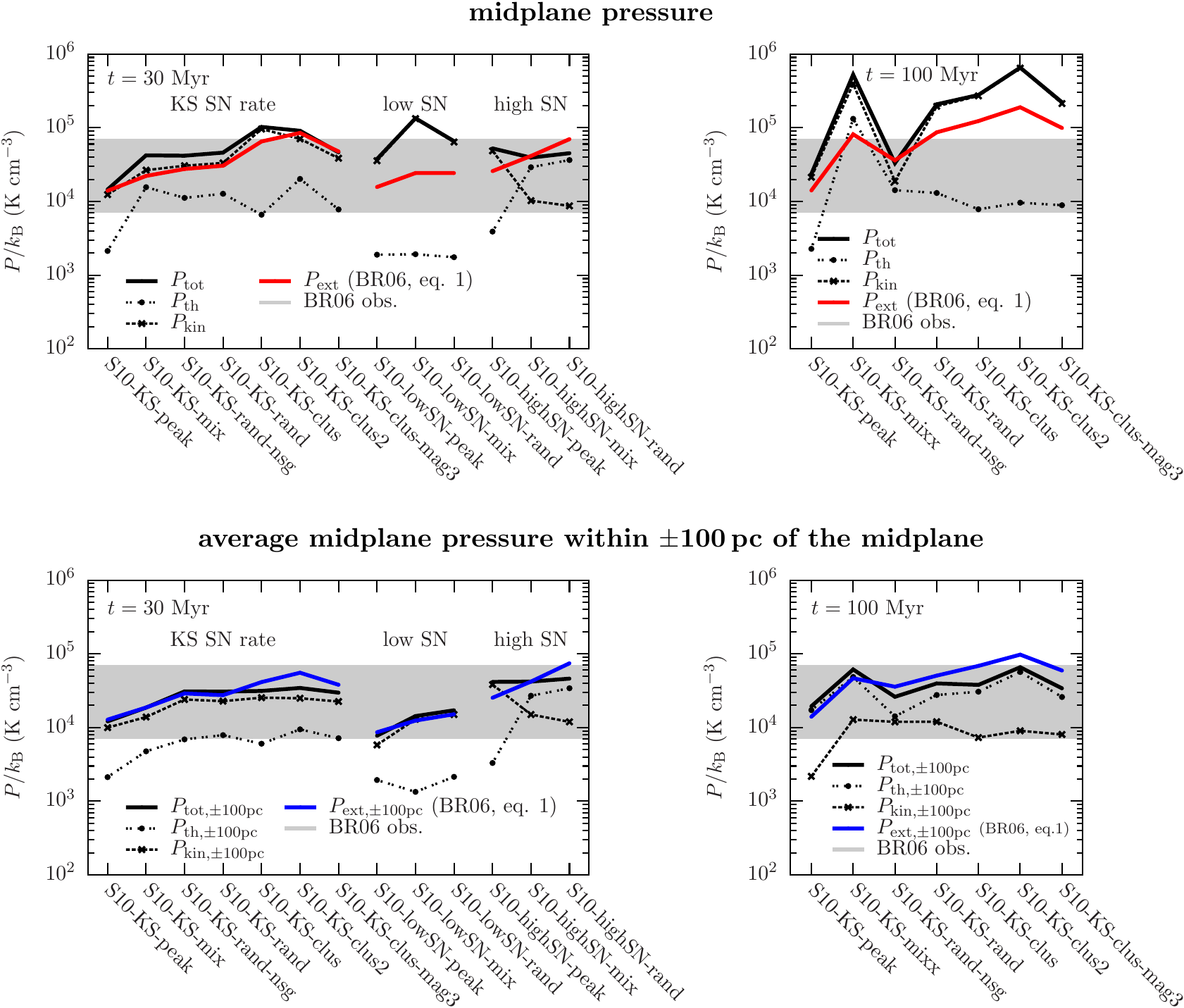}
    \caption{Midplane pressures for all simulations at different times. The upper plots show the pressures in the midplane at $t=30\,\mathrm{Myr}$ (left) and at $t=100\,\mathrm{Myr}$ (right). The lower plots show averaged pressures in the $z$-range $[-100\,\mathrm{pc};\,100\,\mathrm{pc}]$. The black lines denote values directly taken from the simulation data, split into kinetic and thermal components. The coloured lines are the ''interstellar pressure'' computed using Eq.~(\ref{eq:BR06-interstellar-pressure}) \citep[][Eq. (1)]{BlitzRosolowsky2006}. The grey area indicates the range of observed midplane pressures by \citet{BlitzRosolowsky2006}. In almost all cases the kinetic pressure dominates over the thermal one. Using the midplane simulation data yields higher values than the estimated ''interstellar pressure''. Averaging the region within $\pm100\,\mathrm{pc}$ reduces the relative difference below a factor of two.}
    \label{fig:midplane-pressure-split}
  \end{minipage}
\end{figure*}
%%%%%%%%%%%%%%%%%%%%%%%%%%%%%%%%%%%%%%%%%%%%%%%%

A more quantitative analysis is given in Fig.~\ref{fig:midplane-pressure-split}, where we show the mean pressure in a thin slice averaged over the midplane. The left panels show the values at $t=30\,\mathrm{Myr}$ for all simulations, the right panels are at $t=100\,\mathrm{Myr}$. In the top row we use the values at the midplane, in the bottom plots we average the numbers in the $z$ range $\pm100\,\mathrm{pc}$. The black lines are the values directly obtained from the simulation box. The grey area indicates the midplane pressures inferred from observations by \citet{BlitzRosolowsky2006}. For this comparison we have to compute the ''interstellar pressure'', $P_\mathrm{ext}$, in the same way as \citet{BlitzRosolowsky2006}, cf. their equation~(1),
\begin{equation}
  \label{eq:BR06-interstellar-pressure}
  P_\mathrm{ext} = (2G)^{1/2}\Sigma_\mathrm{g}\sigma_\mathrm{g}\left[\rho_\star^{1/2} + \left(\frac{\pi}{4}\rho_\mathrm{g}\right)^{1/2}\right],
\end{equation}
with $G$ being Newton's constant, $\Sigma_\mathrm{g}$ the surface density of the gas, $\sigma_\mathrm{g}$ the total velocity dispersion of the gas, and $\rho_\star$ and $\rho_\mathrm{g}$ the stellar and gas densities. For the stellar density, $\rho_\star$, we use the same values as for the external potential (see Sec.~\ref{sec:num-methods}). The values for $P_\mathrm{ext}$ are shown with the coloured lines.

The numbers support the visual impression that the kinetic pressure dominates over the thermal one in almost all runs. In the simulations S10-highSN-mix and S10-highSN-rand the disc is already blown apart which explains the low kinetic and high thermal pressures. The total pressures taken directly from the midplane in the simulations are at the upper end of the observed values at $t=30\,\mathrm{Myr}$. At the end of the simulation at $t=100\,\mathrm{Myr}$ almost all numbers clearly exceed the observed range. Computing $P_\mathrm{ext}$ using the gas density and the velocity dispersion yields values which are still high but in agreement with the observations.

Since the observations by \citet{BlitzRosolowsky2006} are only resolved down to $\sim100\,\mathrm{pc}$ it seems reasonable to also compare their pressures to the simulation values based on averaged quantities from $\pm100\,\mathrm{pc}$. The central high values of the kinetic pressure are then attenuated by a factor of up to an order of magnitude. Using the averaged quantities from $\pm100\,\mathrm{pc}$ to compute $P_\mathrm{ext}$ shows a similar agreement of the total pressures taken from the simulations with the observed estimates.

\subsection{Vertical pressure profiles}

%%%%%%%%%%%%%%%%%%%%%%%%%%%%%%%%%%%%%%%%%%%%%%%%
\begin{figure*}
  \begin{minipage}{\textwidth}
    \centering
    \includegraphics[width=\textwidth]{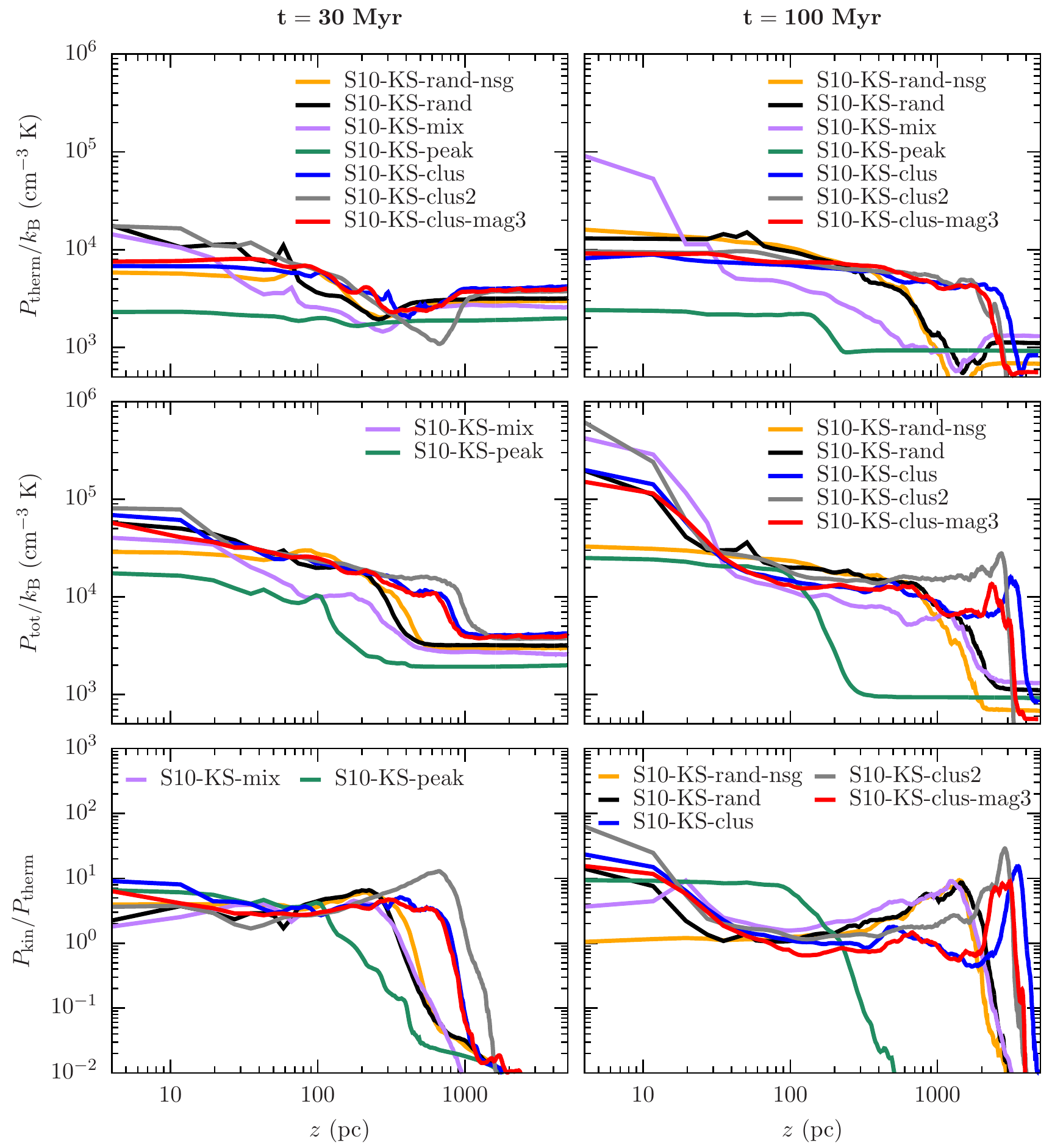}
  \end{minipage}
  \caption{Vertical profiles of the thermal (top) and total pressure (middle) as well as the ratio of kinetic to thermal pressure (bottom) at $t=30\,\mathrm{Myr}$ (left) and $t=100\,\mathrm{Myr}$ (right). The thermal pressure shows moderate variations along the $z$ coordinate at an early stage of the simulation. After $100\,\mathrm{Myr}$ the variations increase up to two orders of magnitude. On the other hand, the total pressure shows an outward gradient early on which becomes stronger over time. At $t=100\,\mathrm{Myr}$ the central pressure differs noticeably. Runs S10-KS-peak and S10-KS-rand-nsg show similar values which are an order of magnitude lower compared to the other runs. At $t=30\,\mathrm{Myr}$ the ratio of kinetic to thermal pressure indicates that the kinetic pressure is $\sim5$ times higher than the thermal one in the disc. Above $z\sim200-1000\,\mathrm{pc}$ the thermal component dominates and the flow becomes sub-sonic.}
  \label{fig:S10-vert-pressure-profiles}
\end{figure*}
%%%%%%%%%%%%%%%%%%%%%%%%%%%%%%%%%%%%%%%%%%%%%%%%

In Fig.~\ref{fig:S10-vert-pressure-profiles} we plot from top to bottom vertical profiles of the thermal ($P_\mathrm{therm}$) and the total pressure,
\begin{equation}
  \label{eq:total-pressure}
  P_\mathrm{tot} = \langle\rho v^2\rangle + \langle P_\mathrm{therm}\rangle + \frac{\langle B^2\rangle}{8\pi},
\end{equation}
as well as their ratio at $t=30\,\mathrm{Myr}$ (left) and $t=100\,\mathrm{Myr}$ (right). For $t\lesssim50\,\mathrm{Myr}$ the thermal pressure shows moderate variations and local fluctuations along the $z$ coordinate. It is not surprising that the peak driving run has the lowest values of $P_\mathrm{therm}$ because the SNe are mostly injected as momentum rather than as thermal energy. The thermal pressure in the midplane ranges from $P_\mathrm{therm}/k_\mathrm{B}=2-20\times10^3\,\mathrm{cm}^{-3}\,\mathrm{K}$. At the end of the simulation the thermal pressure at high altitudes decreases by an order of magnitude due to cooling. The central values do not change significantly over time except for run S10-KS-mix, which forms a dense GMC towards the end of the simulation, in which energy is deposited by the fraction of SNe placed in density peaks. In contrast the total pressure changes noticeably. At $t=30\,\mathrm{Myr}$ $P_\mathrm{tot}$ drops by a factor of 10 from the centre to high altitudes ($z\gtrsim0.5\,\mathrm{kpc}$). The ratio of kinetic to thermal pressure ($P_\mathrm{kin}/P_\mathrm{therm}=\mathcal{M}^2$ with the Mach number $\mathcal{M}$) below reveals that the innermost $\sim200-1000\,\mathrm{pc}$ are dominated by kinetic contributions. Above that height the thermal pressure dominates and the flow becomes sub-sonic. Up to that evolutionary stage the difference between the individual runs is not dramatic (except for the height at which the ratio drops). At $t=100\,\mathrm{Myr}$ the total pressure in the midplane increased by a factor of a few up to one order of magnitude except for run S10-KS-rand-nsg. This inner part of the disc up to $z\sim50\,\mathrm{pc}$ is dominated by kinetic pressure and coincides with the $z$ positions of the majority of the SNe. The height from $50-500\,\mathrm{pc}$ is at pressure ratios of order unity. The drop in total pressure at $z>1\,\mathrm{kpc}$ marks the front of the outflowing gas.

\subsection{Outflows and mass loading}
\label{sec:outflows}

%%%%%%%%%%%%%%%%%%%%%%%%%%%%%%%%%%%%%%%%%%%%%%%%
\begin{figure*}
  \begin{minipage}{\textwidth}
    \centering
    \includegraphics[width=0.9\textwidth]{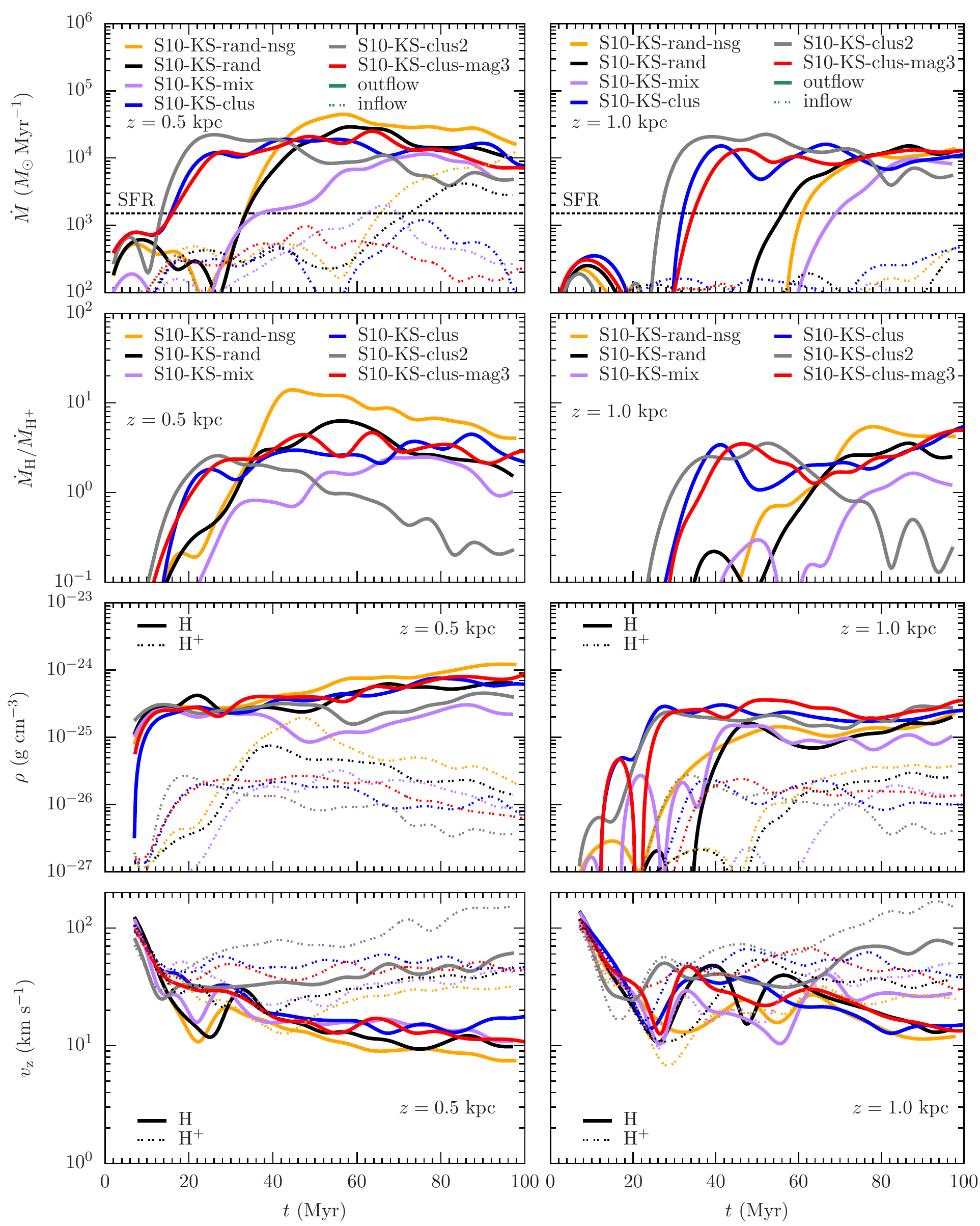}
   \end{minipage}
  \caption{Time evolution of the outflow properties at $0.5\,\mathrm{kpc}$ (left) and $1\,\mathrm{kpc}$ (right). The upper plots show the outflow (lines) and inflow rates (dots) which are generally lower. A coherent outflow is only launched with random or clustered driving and only after $t\sim10-30\,\mathrm{Myr}$ ($0.5\,\mathrm{kpc}$) or $t\sim30-50\,\mathrm{Myr}$ ($1.0\,\mathrm{kpc}$), respectively. Comparing the star formation rate (SFR, see Sec.~\ref{sec:SN-energy-input}) to the outflow rate yields mass loading factors of up to 10 at the end of the simulations. The panel below depicts the composition of the outflow. We note that the outflow does not contain any H$_2$ but mainly consists of atomic hydrogen. For clustered driving at $z=0.5\,\mathrm{kpc}$ the fraction of H is lower in favour of H$^+$. The third row shoes the outflow density, which is remarkably similar for all runs in atomic hydrogen. The outflow velocity in the bottom panels is a factor of a few lower for atomic hydrogen compared to ionized hydrogen.}
  \label{fig:S10-mass-loading-factor}
\end{figure*}
%%%%%%%%%%%%%%%%%%%%%%%%%%%%%%%%%%%%%%%%%%%%%%%%

Because of the short simulation times we cannot discuss galactic outflows and fountain effects in a steady state \citep[see][]{deAvillezBreitschwerdt2004,HillEtAl2012}. However, we can investigate the initial onset of an outflow as a function of driving mode and SN rate. We measure outflow activity as the flux of mass at $z=0.5\,\mathrm{kpc}$ and at $z=1\,\mathrm{kpc}$ and relate this quantity to the star formation rate. This gives the mass loading factor
\begin{equation}
  \eta(z) = \frac{\dot{M}(z)}{\dot{M}_\mathrm{SFR}}.
\end{equation}
Outflowing and inflowing gas is traced separately instead of only tracing the \emph{net} flow. We would like to emphasize that \emph{inflow} measured at height $z$ is only gas that has been pushed out before and now falls back towards the midplane of the disc. We have no external accretion flow onto the disc, i.e. from outside the simulation box. In addition, we analyse the composition of the outflow by splitting up the outflowing mass into the individual hydrogen components.

The previous analysis demonstrates that the peak driving model has a compact disc with no or very little mass being launched out of the disc. In this section we thus discard this model. In Fig.~\ref{fig:S10-mass-loading-factor} we summarize the outflow properties over time. The plots on the left are the values at $z=\pm0.5\,\mathrm{kpc}$, the plots on the right at $z=\pm1.0\,\mathrm{kpc}$. The top row shows the outflow (solid lines) and inflow (dotted lines). The chemical composition is shown as the ratio of atomic hydrogen mass over the mass in H$^+$ in the second row. None of the outflows carries any H$_2$ or CO; they are composed solely of H and H$^+$. This result is possibly affected by resolution. With a minimum cell size of $4\,\mathrm{pc}$ we are unable to resolve typical substructures in dense regions. Higher resolution allows for the formation of denser regions that can form molecular gas more easily. In addition, stable dense structures might also resist the dissolution by hot fast gas more effectively. The third and fourth row present the density and the mass-weighted outflow velocities, separately for atomic hydrogen (solid lines) and ionized hydrogen (dotted lines).

We note that our simulations establish strong outflows after a short time of $\sim10-30\,\mathrm{Myr}$. As we start our calculations with a simple isothermal disc without any substructure and velocity fluctuations, we need to wait for the SNe to insert enough energy and momentum before measurable outflows are launched. Clustered SNe start launching outflows earlier than individual SNe. The inflow is smaller over the entire simulation time such that we have net outflow with a mass loading factor of up to $10$. Only for random driving does the inflow increase towards the end with inflow rates larger than the star formation rate. At a height of $z=1\,\mathrm{kpc}$ we notice a similar behaviour with a delay of $\sim20\,\mathrm{Myr}$ corresponding to a net outflow velocity of $25\,\mathrm{km}\,\mathrm{s}^{-1}$.

The ratio of H to H$^+$ in Fig.~\ref{fig:S10-mass-loading-factor} indicates that the composition is dominated by atomic hydrogen for all runs except S10-KS-clus2 at the end of the simulation. Strong SN clustering generally lowers the amount of atomic hydrogen. The fact that S10-KS-clus2 without any SNe of type~Ia shows the lowest ratio suggests that the SN positioning/clustering in the disc has a stronger impact on the composition of the outflow than the inclusion of individual type~Ia SNe with a six times larger distribution in $z$. At $z=1\,\mathrm{kpc}$ the correlation with the driving mode is less pronounced.

We compute the mean outflow density assuming that within one computational cell the chemical mass fractions equal the corresponding volume fractions. At $z=0.5\,\mathrm{kpc}$ there is only little scatter in the density of atomic hydrogen with $\rho\sim2-10\times10^{-25}\,\mathrm{g}\,\mathrm{cm}^{-3}$ for all simulations, slowly rising over time. The values at $1\,\mathrm{kpc}$ are very stable with similar densities. For ionized hydrogen we measure about one to two orders of magnitude lower numbers with a larger temporal scatter.

%%%%%%%%%%%%%%%
\begin{figure}
  \centering
  \includegraphics[width=8cm]{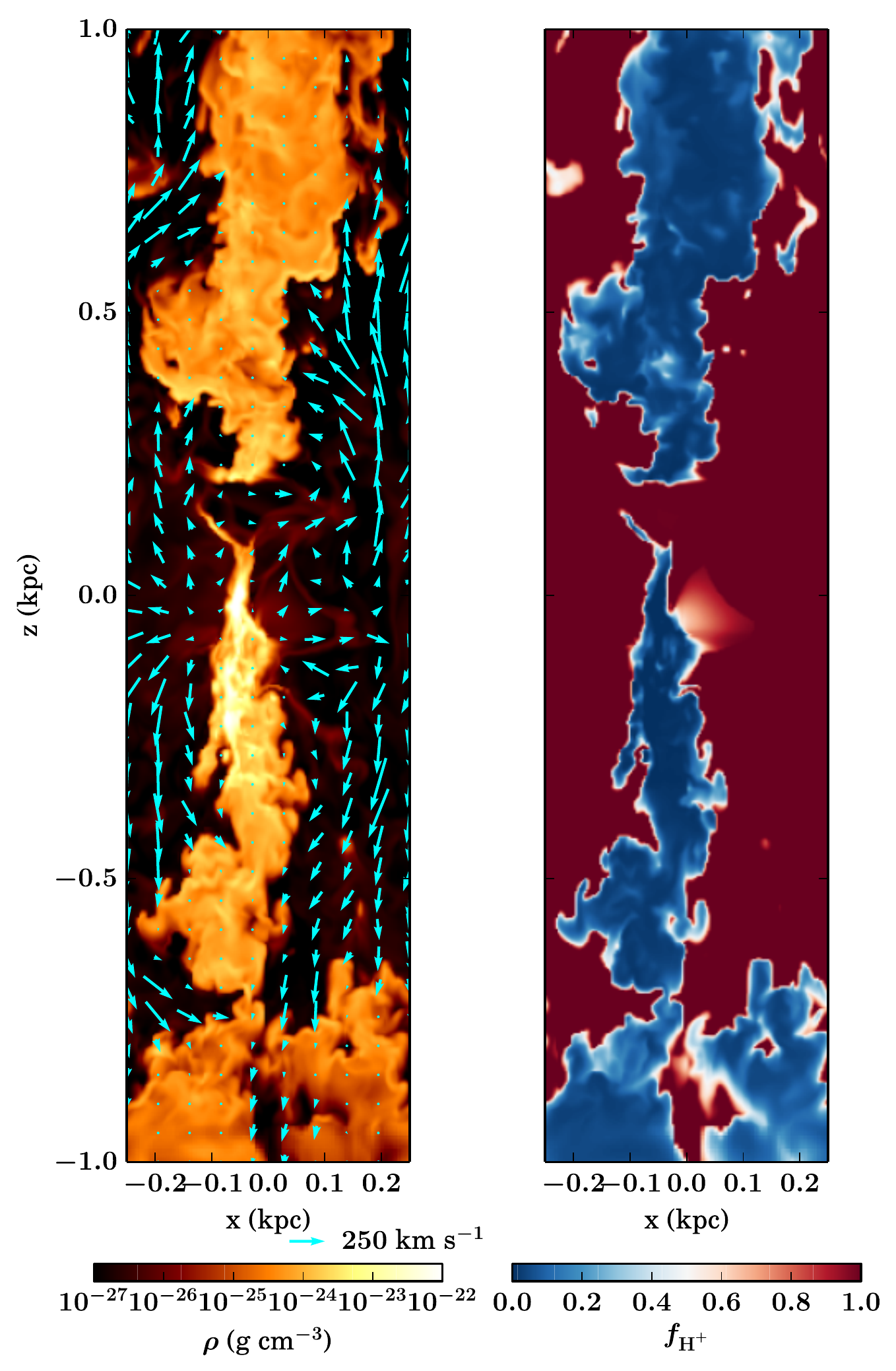}
  \caption{Cuts through the box of the gas density with outflow velocity vectors (left) and ionization fraction (right) for simulation S10-KS-rand at $t=100\,\mathrm{Myr}$. At $|z|=0.5\,\mathrm{kpc}$ the outflow follows the structure of a collimated low-density chimney, at $|z|=1\,\mathrm{kpc}$ the flow converts into more turbulent motions. The low-density ionized gas reaches velocities of a few hundred $\mathrm{km}\,\mathrm{s}^{-1}$ whereas the dense gas moves at velocities of the order of $10\,\mathrm{km}\,\mathrm{s}^{-1}$.}
  \label{fig:outflow-vectors}
\end{figure}
%%%%%%%%%%%%%%%

The panel at the bottom depicts the mean outflow velocities. At $z=0.5\,\mathrm{kpc}$ there is a difference between the mass-weighted H velocities of about $10\,\mathrm{km}\,\mathrm{s}^{-1}$ and the H$^+$ velocities which are three times as high at the end of the simulation. S10-KS-clus2 shows the largest H velocities of $\sim50\,\mathrm{km}\,\mathrm{s}^{-1}$ at the end of the simulation. At $z=1\,\mathrm{kpc}$ the scatter in the velocities is very high with a less pronounced distinction between H and H$^+$ due to turbulent instabilities and mixing. This behaviour is illustrated in Fig.~\ref{fig:outflow-vectors} where we show the density with velocity vectors (left) and the H$^+$ fraction of the gas (right) in cuts through the box for simulation S10-KS-rand at $t=100\,\mathrm{Myr}$. We find high velocities of up to several hundred $\mathrm{km}\,\mathrm{s}^{-1}$ in the low-density gas that starts as a collimated chimney and mixes with the high-density, low-velocity gas at $z\gtrsim0.5-1\,\mathrm{kpc}$.

%%%%%%%%%%%%%%%%%%%%%%%%%%%%%%%%%%%%%%
\begin{figure}
  \includegraphics[width=8cm]{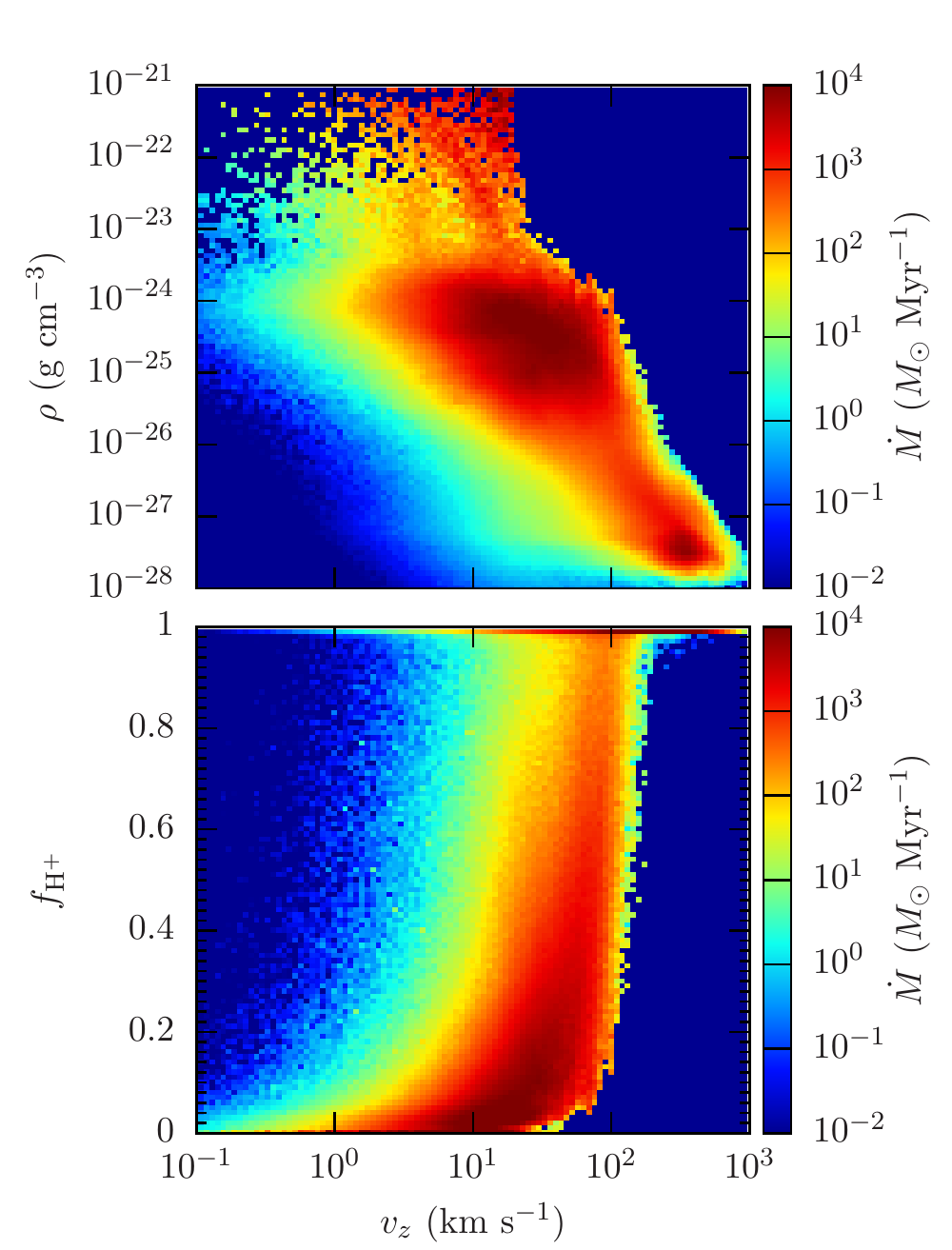}
  \caption{Two-dimensional histograms for $v_z-\rho$ (top) and $v_z-f_\mathrm{H^+}$ (bottom) for the outflowing gas for simulation S10-KS-clus-mag3 at $t=100\,\mathrm{Myr}$. Colour-coded is the outflow rate. There is almost a bimodal distribution in the outflow rate (upper plot). The bulk of the outflow has velocities of a few tens of $\mathrm{km}\,\mathrm{s}^{-1}$ at a mean density of $\rho\sim10^{-24}\,\mathrm{g}\,\mathrm{cm}^{-3}$. We note that the high-velocity gas ($v_z>100\,\mathrm{km}\,\mathrm{s}^{-1}$) is fully ionized and mostly at very low densities. However, this high-velocity tail contributes significantly to the total outflow rate. Below $v_z\lesssim100\,\mathrm{km}\,\mathrm{s}^{-1}$ the fraction of ionized gas shows a wide spread.}
  \label{fig:outflow-phase-plots}
\end{figure}
%%%%%%%%%%%%%%%%%%%%%%%%%%%%%%%%%%%%%%

In Fig.~\ref{fig:outflow-phase-plots} we illustrate how the density and the composition of the outflow vary with the vertical velocity. Plotted are two-dimensional histograms for $v_z-\rho$ (top) and $v_z-f_\mathrm{H^+}$ (bottom) for the outflowing gas for simulation S10-KS-clus-mag3 at $t=100\,\mathrm{Myr}$. The colour-coded outflow rate indicates an almost bimodal distribution, with the bulk of the outflow moving at $10-100\,\mathrm{km}\,\mathrm{s}^{-1}$ at a mean density of $\rho\sim10^{-25}-10^{-24}\,\mathrm{g}\,\mathrm{cm}^{-3}$. Although the high-velocity gas ($v_z>100\,\mathrm{km}\,\mathrm{s}^{-1}$) is at low densities it contributes significantly to the total outflow rate.

Given the outflow velocities and densities we can estimate the amount of gas that can escape from the disc for our choice of the external gravitational potential neglecting further outward acceleration. 
%In order to leave the simulation box at $z=\pm5\,\mathrm{kpc}$ the velocity at $z=0$ needs to be $v_z>90\,\mathrm{km}\,\mathrm{s}^{-1}$. The total gas mass that can leave the computational domain ranges from $M_\mathrm{esc,box}\sim0.8-7\times10^4\,M_\odot$ from mixed over random to clustered SN driving, which corresponds to $0.3-2.7$ percent of the disc mass.
In order to escape from the Milky Way the gas at the solar radius must have velocities above $v_\mathrm{esc}(10\,\mathrm{kpc})\sim500\,\mathrm{km}\,\mathrm{s}^{-1}$ \citep{XueEtAl2008}. For $v_\mathrm{esc}=500\,\mathrm{km}\,\mathrm{s}^{-1}$ the fraction of escaping mass is only $2-8\times10^{-5}$, so basically no gas is expected to escape from Milky Way gravitational potential based on the $z$ velocities we measure at the end of the simulation. The fraction of the volume with outward velocities above $500\,\mathrm{km}\,\mathrm{s}^{-1}$ is significantly larger and ranges from $1-30\%$ based on a total volume of $0.5\times0.5\times4\,\mathrm{kpc}^3$ excluding the regions of unperturbed pristine gas far above and below the disc.

Determining the mass loading factor, $\eta$, from observations is very difficult and involves many assumptions about the composition of the gas and the structure of outflows \citep[see discussion on clumpy versus smooth outflows in the appendix of][]{MartinEtAl2013}. Observed outflows indicate mass loading factors of order unity \citep{ChenEtAl2010,MartinEtAl2012,NewmanEtAl2012bShort}. Simulations of galaxies also agree with those numbers \citep[see, e.g.][]{OppenheimerDave2008,HopkinsQuataertMurray2012}. At $z=0.5\,\mathrm{kpc}$ we reach values of up to $\eta=10$ which are higher than the observational estimates. However, we only simulate the onset of fountains and winds for time-scales shorter than needed for typical galactic fountains. Gas that eventually falls back towards the midplane can perceptibly decrease the net mass loading factor. A conclusive statement concerning the connection between local mass loading factors at $z=0.5-1\,\mathrm{kpc}$ and escaping gas ($z\gg10\,\mathrm{kpc}$) cannot be drawn based on our simulations.

\citet{CreaseyTheunsBower2013} investigate the mass loading factor in detail as a function of disc surface density and the ratio of gas to dark matter. Their surface densities are overall higher than our value of $\Sigma=10\,M_\odot\,\mathrm{pc}^{-2}$, which does not allow for a direct comparison. In addition, they impose a temperature floor at $10^4\,\mathrm{K}$, which might measurably change the structures in the midplane where the SNe explode and thus the outflow properties. However, extrapolating their curves of the mass loading factor down to our surface density of $\Sigma=10\,M_\odot\,\mathrm{pc}^{-2}$ yields very similar results.

\section{Higher and lower SN rates}

In all runs with lower SN rate (three times lower than the KS SN rate) the gas collapses into a very thin sheet. The energy input from the SNe is not able to support the disc against the gravitational attraction. Molecular gas forms very efficiently. As in the KS simulations, the peak and mixed SN models do not drive any outflow. Randomly placed SNe (S10-lowSN-rand) are able to temporarily push gas out to $z\sim700\,\mathrm{pc}$, but at very low speeds. Only in the last few Myr of the simulation time a chimney establishes and a weak outflow is launched.

In the case of high SN rates (three times as high as the KS SN rate) the evolution of the disc depends even more strongly on the driving mechanism. For random (S10-highSN-rand) and mixed (S10-highSN-mix) SN positions the kinetic pressure is high enough to completely evacuate the midplane region after $\sim20-30\,\mathrm{Myr}$. As a factor of a few higher SN rate is not implausible for real systems, the temporal effects of starburst events and the resulting cavities in the disc midplane might not be unphysical, i.e. the existence of a temporally evacuated central region of the disc does not mark the model as unphysical. However, at this point the constant SN rate with positions in the evacuated region is not justified any more. In simulation S10-highSN-mix the gas remains in the region around $\pm0.5\mathrm{kpc}$ with a balanced force from SN driving and gravity. Simulation S10-highSN-rand does not show a turnover of the gas within the simulated time. In the case of SNe in density peaks (S10-highSN-peak) the energy input is strong enough to efficiently mix the gas on small scales and thus prevents the formation of large density contrasts, so the disc has very smooth density distribution extending from $-0.5-0.5\,\mathrm{kpc}$. Similar to the run with the KS SN rate no outflow is launched.

%%%%%%%%%%%%%%%%%%%%%%%%%%
\begin{figure}
  \centering
  \includegraphics[width=8cm]{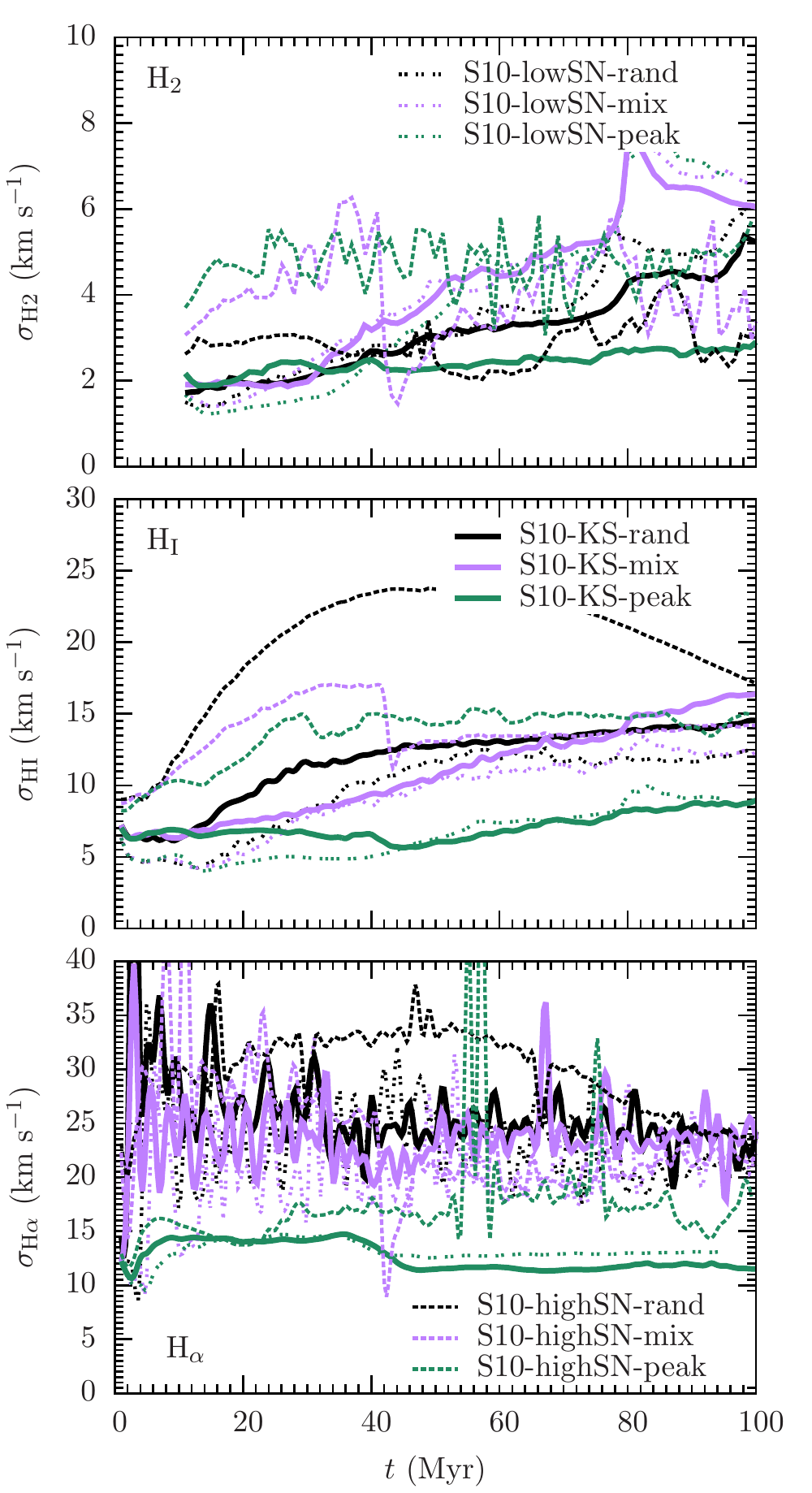}
  \caption{Time evolution of the velocity dispersions for H$_2$ (top) as well as the observational estimates for H$_\textsc{i}$ (middle) and H$_\alpha$ (top) for different SN rates and driving modes. A factor of three in the SN rate can double the velocity dispersion. Peak driving results in lower $\sigma$ compared to mixed and random driving. At a later stage of the simulations the impact of the driving mode can be larger than a three times higher or lower SN rate.}
  \label{fig:velocity-dispersion-SNrate-comp}
\end{figure}
%%%%%%%%%%%%%%%%%%%%%%%%%%

In Fig.~\ref{fig:velocity-dispersion-SNrate-comp} we compare the velocity dispersions for peak, mixed, and random SN driving for low, KS, and high SN rates. The upper panel depicts $\sigma_\mathrm{H2}$, the middle and lower panel present the observational estimates for H$_\textsc{i}$ and H$_\alpha$. For $\sigma_\mathrm{H2}$ we notice an early trend with SN rate which becomes weaker over time and is lost after $\sim30\,\mathrm{Myr}$ when the first clouds and voids in the disc have formed (see Fig.~\ref{fig:coldens-plane-structure-evolution}). The values span a range from $2-6\,\mathrm{km}\,\mathrm{s}^{-1}$. The evolution of $\sigma_\mathrm{HI}$ reveals an overall dependence on the SN rate but towards the end an even stronger dependence on the driving mechanism. SNe in density peaks lead to the lowest values followed by models with mixed and random SN positioning. In $\sigma_\mathrm{H\alpha}$ the temporal scatter is very large but overall the velocity dispersions are determined to first order by the driving mechanism with the same ordering as for $\sigma_\mathrm{HI}$.

%%%%%%%%%%%%%%%%%%%%%%%%%%%%
\section{Potential caveats}%
%%%%%%%%%%%%%%%%%%%%%%%%%%%%
\label{sec:discussion}

\subsection{Dynamical equilibrium}

Previous studies show that the formation of molecular gas in colliding flows of atomic gas takes about $10\,\mathrm{Myr}$ \citep{ClarkEtAl2012}. Two molecular clouds initially at rest with a mass of $10^5\,M_\odot$ at a distance of $100\,\mathrm{pc}$ need about $40\,\mathrm{Myr}$ to merge due to their mutual gravitational attraction. This adds up to time-scales of around $50\,\mathrm{Myr}$ for the formation of a GMC. In a turbulent environment the formation and merging of clouds is likely to occur on shorter time-scales, which has also been suggested by observations \citep{TamburroEtAl2008, FukuiEtAl2015} finding time-scales as small as $\sim10^6\,\mathrm{yr}$.

Our simulated time-scale of $100\,\mathrm{Myr}$ is thus expected to cover the formation of molecular clouds, and indeed we observe this process in all simulations (see Fig.~\ref{fig:coldens-plane-structure-evolution}, where we show the total column density for simulation S10-KS-rand over time). However, we see no dispersing or dissolving clouds. We conclude that SNe alone cannot destroy massive molecular clouds, even if they are all placed in density peaks, i.e. in the centre of molecular clouds. A dynamical equilibrium in a sense of a constant number of clouds or a constant average cloud mass is not established in our simulation.

This leads to the conclusion that either the assumption of a constant star formation and SN rate is not appropriate on the scales under consideration or that we neglect important physical processes in our model such as feedback.

\subsection{Assumption of a Constant SFR}
On large scales, i.e. when considering a significant fraction of the Galactic disc with a statistically relevant number of star-forming regions, the assumption of an averaged constant SN rate that scales directly with the surface density is observationally well justified. This is known as the Kennicutt-Schmidt (KS) relation (\citealt{Schmidt1959}, \citealt{Kennicutt1998}, \citealt{BigielEtAl2008}). Locally, the KS relation breaks down, which is expected because individual star-forming complexes together with their temporal evolution need to be considered \citep[see also][]{KruijssenLongmore2014}. Time-dependent SN rates that are linked to the formation of molecular gas are likely to prevent the continuous increase of the average cloud mass by significantly increasing the local (spatial and temporal) SN feedback compared to the constant SN rate based on the KS relation. In simulations with very simplified physics this problem might not occur at all because of the missing stable dense gas structures that serve as star forming regions. Indeed, it has been shown that a simplified thermodynamic model and no self-gravity does statistically result in a turbulent ISM with reasonable fractions of hot and cold gas \citep[see, e.g.][]{deAvillezBreitschwerdt2004}. However, no long-lived stable structures form in those simplified setups. The small-scale transient features can still be treated and efficiently stirred with a globally oriented star formation and SN prescription to give reasonable values in terms of dynamics (e.g. velocity dispersion) and thermodynamics (e.g. volume-filling fractions).

\subsection{Stellar feedback}
It is likely that additional stellar feedback processes provide additional support against gravitational collapse, e.g. protostellar outflows, radiation feedback, and stellar winds. Early on during the formation process of massive stars protostellar outflows can significantly stir the gas in molecular clouds \citep{SeifriedEtAl2012,FederrathEtAl2014,OffnerAcre2014,PetersEtAl2014} providing mechanical energy against further collapse of molecular clouds. Radiation feedback has long been suggested as an important heating source that can suppress gravitational collapse and expel gas from clouds \citep{Stromgren1939, Kahn1954, OortSpitzer1955}. Numerical studies support this paradigm and show efficient heating of the surroundings by OB stars \citep{DaleEtAl2005,PetersEtAl2010a,ArthurEtAl2011,WalchEtAl2012b,DaleEtAl2014,GeenEtAl2015}. Recent work by \citet{BonebergEtAl2015} shows that radiation feedback can even drive turbulence in molecular clouds. Although being generally less efficient than UV photoheating, direct radiation pressure is expected to drive dynamical feedback \citep{KrumholzMatzner2009,MurrayQuataertThompson2010,KrumholzThompson2012,SalesEtAl2014}. Stellar winds might equally well counteract gravitational attraction due to the formation of hot low-density bubbles, found by theoretical estimates \citep{Avedisova1972,CastorMcCrayWeaver1975,WeaverEtAl1977} as well as recent numerical simulations \citep{RogersPittard2013,DaleEtAl2014}.

\subsection{Global dynamics}

Our simulations also imply certain assumptions of the global dynamical properties of the galaxy under consideration. Most notably, we do not include the shear motions that are characteristics of the gas flow in typical disc galaxies. In order to determine the potential impact of galactic rotation we evaluate the Rossby number,
\begin{equation}
  \epsilon_\mathrm{R}=\frac{v_\mathrm{gas}}{\Omega L},
\end{equation}
which is defined as the ratio of inertial to Coriolis force. Here, $v_\mathrm{gas}$ is the velocity of the gas, $\Omega=v_\mathrm{rot}/(2\pi R)$ is the angular frequency with which the stratified box would rotate around the centre of the galaxy, $v_\mathrm{rot}$ the tangential velocity, $R$ the distance from the galactic centre, and $L$ is the size of the box in the rotation plane (dimension in $x$ and $y$ in our case). Thus small numbers indicate that the system is affected by the Coriolis force. For our box with $L=0.5\,\mathrm{kpc}$ and typical numbers of $v_\mathrm{gas}\sim10\,\mathrm{km}\,\mathrm{s}^{-1}$, $v_\mathrm{rot}=200\,\mathrm{km}\,\mathrm{s}^{-1}$ \citep{ReidEtAl1999}, $R=8\,\mathrm{kpc}$ ($\Omega\approx8\times10^{-17}\,\mathrm{s}^{-1}$) the Rossby number is $\epsilon_\mathrm{R}\approx5$. The gas in our box reaches velocities a factor of a few higher and most of the structures we are concentrating on are smaller than the size of the total box. Both corrections increase the Rossby number, so our simulations are not expected to be influenced by rotation. However, as shear might still be important for some parts of the dynamics in the ISM, we are going to implement shearing box conditions for future projects.

\section{Summary and conclusions}

With the SILCC simulations we aim at better understanding the dynamical evolution of the ISM and the life cycle of molecular clouds in star forming disc galaxies like the Milky Way. We simulate the evolution of a part of a galactic disc (a stratified box of size $0.5\times0.5\times\pm5\,\mathrm{kpc}$ with a gas surface density of $\Sigma=10\,M_\odot\,\mathrm{pc}^{-2}$) on a $\sim 100\,\mathrm{Myr}$ time-scale including an external stellar disc potential and self-gravity. Gas is heated by the background interstellar radiation and can cool based on a chemical network incorporating H$^+$, H, H$_2$, CO and C$^+$, which also allows us to follow the formation of molecular gas. We include shielding of the gas and stellar feedback in the form of SNe. The SN rate is based on the Kennicutt-Schmidt (KS) relation assuming a universal initial mass function. We cover the scatter in the KS relation by varying the SN rate by a factor $3$ and $1/3$. The positioning of the SNe is varied between random driving (random locations), peak driving, where the SNe are placed in the density peaks, mixed driving (with equal random and peak contributions), and clustered driving, where a fraction of the SNe are spatially clustered following the available observational constraints. We also test the effect of magnetic fields with an initial field strength of $B=3\,\mu\mathrm{G}$ oriented in the plane of the disc. Some simulations form large molecular cloud complexes that become Jeans unstable and undergo collapse to form stars. At this time the calculation is stopped even if $t<100\,\mathrm{Myr}$. 

Our main results can be summarized as follows:

\begin{itemize} 
\item The positioning of SN explosions plays a major role in determining the dynamics in the ISM. SNe placed at random positions lead to a higher gas velocity dispersion than SNe exploding in density peaks, where much of the input energy can be dissipated efficiently. The effect of SN positioning might easily dominate over variations in the SN rate. Comparing velocity dispersions based on chemical composition (H$_2$), observational estimates (H$_\textsc{i}$, H$_\alpha$) and temperature regimes ($8000\,\mathrm{K};\,3\times10^5\,\mathrm{K}$, $300\,\mathrm{K};\,8000\,\mathrm{K}$, $<300\,\mathrm{K}$) shows good agreement with observations and previous ISM studies ($\sigma_\mathrm{HI}\sim0.8\sigma_{300-8000\,\mathrm{K}}\sim10-20\,\mathrm{km}\,\mathrm{s}^{-1}$, $\sigma_\mathrm{H\alpha}\sim0.6\sigma_{8000-3\times10^5\,\mathrm{K}}\sim20-30\,\mathrm{km}\,\mathrm{s}^{-1}$). We find almost constant ratios of the following velocity dispersions over time: $\sigma_\mathrm{HI}/\sigma_\mathrm{H\alpha}\approx0.6$, $\sigma_\mathrm{HI}/\sigma_\mathrm{H2}\approx3-4$, $\sigma_\mathrm{H\alpha}/\sigma_{8000-3\times10^5\,\mathrm{K}}\approx0.6$, $\sigma_\mathrm{HI}/\sigma_{300-8000\,\mathrm{K}}\approx0.8$, and $\sigma_\mathrm{H2}/\sigma_{<300\,\mathrm{K}}\approx0.4$.

\item Randomly placed SNe (both individual SNe and randomly placed clusters of SNe) drive gaseous outflows after $t\sim10-30\,\mathrm{Myr}$ with mass loading factors of up to 10. In contrast, SNe placed in density peaks do not drive any noticeable outflow. Outflows are launched when the ISM is structured into filaments, clouds and voids and the SN remnants are able to expand into low-density gas. This allows for the formation of low-density chimneys and outflow channels ($\sim200\,\mathrm{pc}$ wide), through which the gas is ejected with velocities of up to a few $100\,\mathrm{km}\,\mathrm{s}^{-1}$. The low-density, high-velocity gas ($v>100\,\mathrm{km}\,\mathrm{s}^{-1}$) is mostly ionized and drags denser atomic hydrogen with it. The bulk of the outflowing mass is dense ($\rho\sim10^{-25}-10^{-24}\,\mathrm{g\,cm}^{-3}$) and slow ($v\sim20-40\,\mathrm{km}\,\mathrm{s}^{-1}$) but there is a high-velocity tail of up to $v\sim500\,\mathrm{km}\,\mathrm{s}^{-1}$ with $\rho\sim10^{-28}-10^{-27}\,\mathrm{g\,cm}^{-3}$, which significantly contributes to the total outflowing mass.

\item The outflows are predominantly composed of atomic hydrogen. If we model only clustered SNe the fraction of ionized hydrogen increases and the outflowing gas is composed of H and H$^+$ in roughly equal proportions. Clustered SNe at random positions start driving outflows earlier than individual SNe at random positions. However, the overall outflow rates are very similar for both driving mechanisms. The primary driver for outflows is the vertical kinetic pressure gradient. In all simulations that generate outflows the kinetic pressure gradient is significantly larger than the thermal counterparts. Due to the finite simulation time of $\sim100\,\mathrm{Myr}$ we are limited to studying the launching processes of outflows. Investigations of galactic fountains require longer time-scales and are beyond the scope of the paper.

\item The ISM evolves without reaching any kind of dynamical equilibrium over the entire simulation time of $100\,\mathrm{Myr}$, i.e. dense filaments and small clumps merge to form GMCs. The clouds typically increase in size and mass. Due to their small volume filling fractions the massive dense clumps of molecular gas cannot be disrupted by randomly placed SNe and the peak SNe dissipate too much energy to efficiently disrupt the clouds. Depending on the positions of the SNe and the SN rate the formation of giant molecular clouds is abetted or retarded but a final coalescence of almost all the cold gas in one or a few giant clouds is inevitable. This is a potential caveat of our simulation setup as we do not attempt to include star formation self-consistently. We do follow the formation of molecular clouds but are unable to simulate their destruction. 

\item Self-gravity is of major importance for developing a realistic structure of the disc. In the absence of self-gravity the amount of molecular gas becomes unrealistically low ($f_\mathrm{H2}\lesssim5\%$ compared to $f_\mathrm{H2}\gtrsim40\%$ including self-gravity) and the velocity dispersion in the molecular gas drops to $\sigma_\mathrm{H2}\sim2\,\mathrm{km}\,\mathrm{s}^{-1}$ ($\sigma_\mathrm{H2}\sim5\,\mathrm{km}\,\mathrm{s}^{-1}$ including self-gravity). The densities and pressures in the midplane are at least an order of magnitude lower compared to simulations including self-gravity and compared to observed estimates. 

\item Magnetic fields do not show a significant dynamical impact: the velocity dispersions, the pressures in the midplane as well as the outflow properties are very similar to the corresponding run without magnetic fields. At the end we find fewer but more massive clouds in the magnetic run, but this is likely to be influenced by the initial conditions for the magnetic field. At the beginning of the simulation the magnetic tension efficiently reduces mixing and the magnetic pressure delays gravitational attraction which manifests in a delay of the formation of dense structures. Once molecular clouds have formed the simulations with and without magnetic fields behave very similarly.  

\end{itemize}

We find the best match to the observational data with randomly placed individual or clustered supernovae. Our calculations demonstrate that a magnetised, self-gravitating disc-like sheet of gas driven at a rate consistent with the global KS relation evolves on relatively short time-scales of $\sim50\,\mathrm{Myr}$ into a 'realistic' multi-phase ISM. Its morphological and kinematic structure as well as its phase fractions and outflow properties agree well with the observational data. Future investigations focusing on a more self-consistent star formation treatment and longer time-scales will be required to establish a potential 'self-regulated' or 'equilibrium' state and estimate the global impact on galaxy evolution.  

\section*{Acknowledgements}
We thank Patrick Hennebelle, Philip Hopkins, Mordecai Mac Low, Eve Ostriker, and Jeremiah Ostriker for inspiring discussions. We also thank the referee for valuable questions and suggestions that helped to improve the manuscript. The SILCC team thanks the Gauss Center for Supercomputing (http://www.gauss-centre.eu) and the Leibniz-Rechenzentrum Garching (www.lrz.de) for the significant amount of computer time for this project and their user support. We thank Christian Karch for the program package \textsc{fy} and the community of the \textsc{yt-project} for the \textsc{yt} analysis package \citep{TurkEtAl2011}, which we use to plot and analyse most of the data.
PG, SW, TN, AG, SCOG, RSK, and CB acknowledge support from the DFG Priority Program 1573 {\em Physics of the Interstellar Medium}.
SW acknowledges the support of the Bonn-Cologne Graduate School, which is funded through the Excellence Initiative.
TN acknowledges support from the DFG cluster of excellence \emph{Origin and Structure of the Universe}.
TP acknowledges financial support through a Forschungskredit of the University of Z\"{u}rich, grant no. FK-13-112.
RW acknowledges support by the Czech Science Foundation grant 209/12/1795 and by the project RVO:67985815 of the Academy of Sciences of the Czech Republic.
RSK, SCOG, and CB thank the DFG for funding via the SFB 881 The Milky Way System (subprojects B1, B2, and B8). RSK furthermore acknowledges support from the European Research Council under the European Community’s Seventh Framework Programme (FP7/2007-2013) via the ERC Advanced Grant STARLIGHT (project number 339177).
The software used in this work was developed in part by the DOE NNSA ASC- and DOE Office of Science ASCR-supported Flash Center for Computational Science at the University of Chicago.

%\section*{Data Dissemination}
%The simulation data presented in this paper are partially available for download at \texttt{http://www.girichidis.de/00-SILCC} including the raw simulation data as well as extracted data fields for comparison with observations. Follow-up simulations will also be published there under different data releases.

\bibliographystyle{mn2e}
\bibliography{astro.bib}

\end{document}